\documentclass[aps,prd,twocolumn,superscriptaddress]{revtex4}
\usepackage{float}
\usepackage{graphicx} 
\usepackage{color} 
\usepackage{natbib}
\usepackage{amsmath,amssymb,amsfonts}
\def\prl{Phys. Rev. Lett.}
\def\prd{Phys. Rev. D}
\def\cqg{Class. Quantum Grav.}

\def\apj{Astrophys. J.}
\def\apjs{Astrophys. J. Suppl.}
\def\apjl{Astrophys. J. Lett.}
\def\aap{Astronomy and Astrophysics}
\def\pr{Phys. Rev.}

\def\mnras{Mon. Not. R. Astron. Soc.}
\def\jqsrt{J. Quant. Spectrosc. Radiative Transfer}

\def\Prad{{\mathcal P}}
\def\rhorad{\bar \rho}
\def\jrad{\bar \jmath}
\def\Srad{\bar S}
\def\eq{Eq.}
\def\eqs{Eqs.}

\def\Fst{F}
\def\Fs{{\mathcal F}}
\def\Gst{G}
\def\Gs{{\mathcal G}}
\def\kappaa{\kappa^{\rm abs}}
\def\kappas{\kappa^{\rm sc}}

\def\Dref{\hat {\mathcal D}}
\def\4G{\Gamma}

\begin{document}

\title{Relativistic radiation hydrodynamics in a reference-metric formulation}

\author{Thomas W. Baumgarte}
\affiliation{Department of Physics and Astronomy, Bowdoin College, Brunswick, ME 04011, USA}

\author{Stuart L. Shapiro}
\affiliation{Departments of Physics and Astronomy, University of Illinois at Urbana-Champaign, Urbana, Il 61801}

\begin{abstract}
We adopt a two-moment formalism, together with a reference-metric approach, to express the equations of relativistic radiation hydrodynamics in a form that is well-suited for numerical implementations in curvilinear coordinates.  We illustrate the approach by employing a gray opacity in an optically thick medium.   As numerical demonstrations we present results for two test problems, namely stationary, slab-symmetric solutions in flat spacetimes, including shocks, and heated Oppenheimer-Snyder collapse to a black hole.  For the latter, we carefully analyze the transition from an initial transient to a post-transient phase that is well described by an analytically-known diffusion solution.  We discuss the properties of the numerical solution when rendered in moving-puncture coordinates.
\end{abstract}


\maketitle

\section{Introduction}
\label{sec:intro}

The coincident detection of gravitational and electromagnetic radiation from GW170817 \cite{GW_discovery} has allowed us to observe directly the late inspiral of binary neutron stars together with a short gamma-ray burst (sGRB) and kilonova launched in its aftermath.   The interpretation of these observations requires theoretical models, which can be provided by numerical relativity simulations (see, e.g., \cite{ShiFHKKST17,RadPZB18,RezMW18,RuiST18,GieDU19}, as well as \cite{Pas17,BaiR17} for reviews).    Since a host of different physical processes and phenomena -- including relativistic magnetohydrodynamics, nuclear reactions and radiation transport (both electromagnetic and neutrinos) -- play important roles in the merger of binary neutron stars and the launch of sGRBs and kilonovae, all of these processes also must be accounted for in the numerical simulations. 

While several current codes can handle at least some of these processes, and can evolve the remnant of neutron star mergers for at least several tens of dynamical timescales, i.e.~tens of milliseconds, the complete modeling of secular processes requires even longer evolution times, posing a formidable challenge for most codes (see, e.g., \cite{RuiLPS16,RadPBZ18,CioKKG19,DePFFLPS20,NedBRDEPSSL20}).  On the other hand, the radiation will propagate radially at large distance, where it is measured, and the remnant will rather quickly settle down into an approximately axisymmetric configuration.  These are just some motivations for considering algorithms in curvilinear coordinates, which can take optimal advantage of such symmetries, be they exact or approximate.

One disadvantage of curvilinear coordinates is the appearance of coordinate singularities.  It turns out, however, that these do not affect the stability of suitable evolution schemes as long as all singular terms are handled analytically.  The latter can be accomplished with the help of a reference-metric formulation (see \cite{BonGGN04,ShiUF04,Bro09,Gou12,MonC12}) together with a proper rescaling of all tensorial quantities.  Such an approach was first demonstrated for Einstein's equations in spherical polar coordinates in full 3+1 dimensions by \cite{BauMCM13}, and very similar methods have now been implemented in the {\tt Einstein Toolkit} (also in spherical polar coordinates \cite{MewZCREB18}), the {\tt NRPy++} code (in more general classes of curvilinear coordinates \cite{RucEB18}), as well as the {\tt SpEC} code (in cylindrical coordinates \cite{Jesetal20}).  

When coupling matter fields to Einstein's equations in this approach, it is advantageous to cast these matter fields in a reference-metric formulation as well.  This has been demonstrated for hydrodynamics in \cite{MonBM14} (hereafter MBM, see also \cite{BauMM15}), magnetohydrodynamics (see \cite{MewZCBEAC20}), as well as electrodynamics (see \cite{BauGH19}), but not yet for radiation hydrodynamics -- which is the subject of this paper.

An exact description of radiation transfer entails solving the Boltzmann equation for the specific (energy-dependent) radiation intensity (see, e.g., \cite{Lin66,MihWM84,BauS10}), which, without any approximation or simplifying assumptions, is well beyond the reach of current numerical codes.   As an approximation, local effects of radiative cooling can be estimated with a leakage scheme (see, e.g.,~\cite{RufJS96,BauJKST96,RosL03,SekKKS11,GalKRF13,PerCK16,RadGLROR16,RadPHFBR18,GizORPCN19}).  Radiation transport can be approximated by evolving the lowest angular moments of the intensity only, and expressing higher-order moments with the help of suitable closure relations (see \cite{Tho81}).  In flux-limited diffusion schemes, only the zeroth-order moment (the radiation energy density) is evolved (see, e.g., \cite{Pom81,BurL86,Sha89,BauST96,RahJJ19,Bruetal20} and references therein).  In a two-moment (so-called M1) scheme, the first-order moment (the radiation momentum density, or flux) is evolved together with the zeroth-order moment (e.g.~\cite{RezM94,FarLLS08,MueJD10,ShiKSS11,CarEM13,SadNTZ13,WanSNKKS14,JusOJ15,Oco15,KurTK16,FouDKNPS18,SkiDBRV19,VinFDHKPS20,WeiOR20}).  In general, the moments depend on energy in addition to location and time, but in so-called ``gray" treatments this dependence is suppressed by integrating over the energy.




In this paper we retrace the derivation of such a gray, two-moment formalism using a reference-metric framework, and present numerical examples.  Specifically, we follow the treatment of \cite{FarLLS08}, hereafter FLLS, in Section \ref{sec:eqs}, focussing on the optically-thick regime, but adopt a reference-metric formalism in order to bring the equations into a form that is suitable for implementation in curvilinear coordinates.  Unlike in some previous treatments we also use a systematic 3+1 decomposition of all tensorial quantities, thereby avoiding the potential for confusion between tensors of different types.  In Section \ref{sec:numerics} we demonstrate the feasibility of solving the equations in spherical polar coordinates by presenting numerical results for two test problems, namely planar radiation hydrodynamics shock problems in flat spacetimes, and Oppenheimer-Snyder collapse to a black hole with radiation.  We carefully analyze the early transient behavior of the radiative quantities for the latter, and compare the subsequent radiation field with an approximate analytical solution derived within the relativistic diffusion approximation \cite{Sha89}.

Throughout this paper we adopt geometrized units with $G = 1 = c$.  We denote spacetime indices with $a, b, c \ldots$ and spatial indices with $i, j, k \ldots$.

\section{Equations}
\label{sec:eqs}

\subsection{Preliminaries}
\label{sec:prelims}

We assume that the spacetime $M$ has been foliated by a family of spatial slices that coincide with level surfaces of a coordinate time $t$.  The spacetime line element can then be written as
\begin{eqnarray} \label{line_element}
  ds^2 & = & g_{ab} dx^a dx^b \nonumber \\
  & = & - \alpha^2 dt^2 + \gamma_{ij} (dx^i + \beta^i dt)(dx^j + \beta^j dt), 
\end{eqnarray}
where $g_{ab}$ is the spacetime metric, $\alpha$ the lapse function, $\beta^i$ the shift vector, and
\begin{equation} \label{gamma_def}
  \gamma_{ab} = g_{ab} + n_a n_b
\end{equation}
the induced spatial metric on the spatial slices.  In the last expression, $n^a$ is the future-pointing normal vector on the spatial slices, which we may express as
\begin{equation} \label{n}
n^a = \alpha^{-1} (1, - \beta^i) \mbox{~~~or~~~} n_a = (- \alpha, 0,0,0).
\end{equation}
  
For applications in curvilinear coordinates it is convenient to introduce a spatial reference metric $\hat \gamma_{ij}$ (see, e.g., \cite{BonGGN04,ShiUF04,Bro09,Gou12}).  In numerical applications it is most natural to choose this reference metric to be the flat metric in whatever coordinate system is used -- in our code, for example, it is taken to be the flat metric in spherical polar coordinates.  This assumption is not necessary, however.  In our treatment below we will assume only that $\hat \gamma_{ij}$ is independent of time (which could also be relaxed, for example for applications in cosmology), and will present an analytical example with a curved reference metric in Section \ref{sec:numerics_OS}.

The Baumgarte-Shapiro-Shibata-Nakamura (BSSN) formulation of Einstein's equations \cite{NakOK87,ShiN95,BauS98}, governing the evolution of the gravitational fields, has been expressed in terms of a reference metric by \cite{Bro09,Gou12}, and implemented numerically, assuming spherical polar coordinates, in \cite{MonC12,BauMCM13}.   In the following we also assume the presence of fluid matter.  The equations governing the fluid follow from conservation of rest mass,
\begin{equation} \label{bar_cons}
\nabla_a (\rho_0 u^a) = 0,
\end{equation}
and conservation of total stress-energy,
\begin{equation} \label{T_cons}
\nabla_a T^{ab} = \nabla_a (T^{ab}_{\rm fluid} + R^{ab} ) = 0.
\end{equation}
Here $\rho_0$ is the fluid's rest-mass density, $u^a$ the fluid's four-velocity, $\nabla_a$ the covariant derivative associated with the spacetime metric $g_{ab}$, and the fluid's stress-energy tensor is given by
\begin{equation}
T^{ab}_{\rm fluid} = \rho_0 h u^a u^b + P g^{ab},
\end{equation}
where $h = 1 + \epsilon + P / \rho_0$ is the specific enthalpy, $\epsilon$ the specific internal energy density, and $P$ the fluid pressure.  In (\ref{T_cons}) we have accounted for the presence of radiation by including the radiation stress-energy tensor $R^{ab}$ introduced in \eq~(\ref{rad_se}) below.  As shown in MBM, the equations of relativistic hydrodynamics can also be rewritten with the help of a reference metric, thereby avoiding some of the numerical problems encountered in curvilinear coordinates, and casting the equations in a framework that meshes well with that for the gravitational field equations.  In Section \ref{sec:eqs_dynamics} we will follow a very similar procedure to rewrite the dynamical equations for the radiation fields.

\subsection{Radiation fields in the fluid frame}
\label{sec:eqs_fluid_frame}

We assume that the radiation stress-energy tensor $R^{ab}$ can be written as
\begin{equation} \label{rad_se}
R^{ab} = E u^a u^b + F^a u^b + u^a F^b + \Prad^{ab}.
\end{equation}
Here $u^a$ is the fluid four-velocity, 
\begin{equation} \label{E}
E = \int d \nu d \Omega I_\nu,
\end{equation}
the radiation energy density as measured by an observer comoving with the fluid, 
\begin{equation} \label{Fa}
\Fst^a = h^a_{~b} \int d\nu d\Omega I_\nu N^b,
\end{equation}
the comoving radiation flux four-vector,
\begin{equation}
\Prad^{ab} = h^a_{~c} h^b_{~d} \int d\nu d\Omega I_\nu N^c N^d
\end{equation}
the comoving radiation stress tensor,
$I_\nu$ is the specific intensity, and
\begin{equation}
h^{a}_{~b} = g^{a}_{~b} + u^a u_b
\end{equation}
the projection operator that projects onto slices orthogonal to the fluid four-velocity.   

To illustrate our approach, we assume for simplicity that the radiation is nearly isotropic everywhere, which is appropriate in media that are optically thick.  In this case, the radiation stress tensor takes the form $\Prad^{ab} = \Prad h^{ab}$, where $\Prad$ is the radiation pressure.  The system of equations may then be closed by adopting an Eddington factor of 1/3, so that  
\begin{equation} \label{closure}
\Prad^{ab} = \Prad h^{ab} = \frac{1}{3} E h^{ab} 
\end{equation}
(see, e.g., \cite{ShiKSS11,CarEM13,SadNTZ13,Fou18,FouDKNPS18,WeiOR20} and references therein for more sophisticated closure schemes).  

In the above integrals, $d \Omega$ is the differential solid angle, $\nu$ is the frequency and $I_\nu = I(x^a; N^i, \nu)$ is the specific intensity of radiation at a location $x^a$, moving in the direction $N^a = p^a / (h\nu)$, all measured in the local Lorentz frame of a fiducial observer.   In the last expression $p^a$ is the photon four-momentum, and $h$ the Planck constant. We also note that $\Fst^a$ is orthogonal to the fluid four-velocity,
\begin{equation} \label{u_dot_F_1}
u_a \Fst^a = 0.
\end{equation}

The dynamical equations governing the radiation can then be written as
\begin{equation} \label{eom}
\nabla_b R^{ab} = - \Gst^a,
\end{equation}
where $\Gst^a$ is the radiation four-force 
\begin{equation} \label{rad_force}
\Gst^a = \rho_0 \kappaa (E - 4 \pi B) u^a + \rho_0 (\kappaa + \kappas) \Fst^a.
\end{equation}
Here $\kappaa$ and $\kappas$ are the (frequency-independent) gray-body absorption and scattering opacities, respectively (see, e.g., FLLS for details).  In (\ref{rad_force}),  the frequency-integrated equilibrium intensity $B(T)$ can be written as
\begin{equation}
4 \pi B = a_R T^4,
\end{equation}
where $T$ is the temperature and $a_R$ a constant.  The value of the latter depends on the type of radiation considered:  for thermal radiation it equals the usual radiation constant $a$; for each flavor of non-degenerate neutrinos or anti-neutrinos it is $(7/16) \,a$; and for all neutrinos and antineutrinos combined it is $(7 N_{\nu} / 8)a$, where $N_\nu$ is the number of neutrino flavors contributing to the thermal radiation (see FLLS); here we assume that $kT \gg m_\nu$, as is the case in most stellar applications.   For situations in which the radiation is in thermal equilibrium with the fluid we have $E = 4 \pi B$, but we will not assume that in general.

The radiation moments $E$ and $F^a$, both describing quantities measured by an observer comoving with the fluid, form the {\em primitive} radiation variables.  Coupling the equations of motion to the evolution of spacetime, it is often advantageous to employ a 3+1 decomposition, and to express the radiation equations in terms of {\em conserved} quantities that are related to quantities measured by normal observers. 

\subsection{Radiation fields in the normal frame}
\label{sec:eqs_normal_frame}

We start by decomposing the tensors appearing in Section \ref{sec:eqs_fluid_frame} into their normal and spatial components.
Using (\ref{gamma_def}), we can write the fluid four-velocity $u^a$, for example, as
\begin{equation}
u^a = g^a_{~b} u^b = \gamma^a_{~b} u^b - n^a n_b u^b. 
\end{equation} 
Defining the Lorentz-factor between normal and fluid observers as
\begin{equation}
W \equiv - n_a u^a = \alpha u^t
\end{equation}
and
\begin{equation} \label{v_def}
v^a \equiv \frac{1}{W} \gamma^a_{~b} u^b = (0, u^i/W + \beta^i/\alpha),
\end{equation}
we may write
\begin{equation} \label{u_decomp}
u^a = W ( v^a + n^a).
\end{equation}
Note that $v^a$ is spatial by construction, $n_a v^a = 0$.  Our definition follows that used in the ``Valencia" formulation of relativistic hydrodynamics, but differs from that used by many other authors, including FLLS,
\begin{equation} \label{v_FLLS}
v^i_{\rm FLLS} \equiv \frac{u^i}{u^t} = \alpha v^i - \beta^i. 
\end{equation}

We similarly decompose the radiation flux four-vector into its normal and spatial parts,
\begin{equation}
\Fs \equiv - n_a \Fst^a = \alpha \Fst^t,~~~~~~~~\Fs^a \equiv \gamma^a_{~b} \Fst^b,
\end{equation}
so that
\begin{equation} \label{F_decomp}
\Fst^a = \Fs^a + \Fs n^a.
\end{equation}
Note that the orthogonality condition (\ref{u_dot_F_1}) can now be expressed as
\begin{equation} \label{u_dot_F_2}
\Fs = v_a \Fs^a.
\end{equation}

Following the same approach for the radiation four-force (\ref{rad_force}) we obtain
\begin{equation}
\Gs \equiv - n_a \Gst^a = \rho_0 \kappaa (E - 4 \pi B) W + \rho_0 (\kappaa + \kappas) \Fs
\end{equation}
and
\begin{equation}
\Gs^a \equiv \gamma^a_{~b} \Gst^b = \rho_0 \kappaa (E - 4 \pi B) W v^a + \rho_0 (\kappaa + \kappas) \Fs^a
\end{equation}

We now decompose the radiation stress-energy tensor (\ref{rad_se}) into purely normal, purely spatial, and mixed components.  Specifically, the purely normal component results in the radiation energy density as observed by a normal observer, 
\begin{align}  \label{rhorad}
\rhorad & \equiv  n_a n_b R^{ab} = \alpha^2 R^{tt} \nonumber \\
& = W^2 E + 2 W \Fs + \Prad(W^2 - 1),
\end{align}
where we have used 
$n_a n_b h^{ab} = W^2 -1$ 
in the last equality.  Adopting the closure relation (\ref{closure}) we obtain
\begin{equation} \label{rho_bar}
\rhorad = \frac{4}{3} W^2 E - \frac{1}{3} E + 2 W \Fs. 
\end{equation}
The mixed normal-spatial components of (\ref{rad_se}) yield the momentum flux as observed by a normal observer,
\begin{align} \label{jrad}
\jrad^a & \equiv  - \gamma^a_{~b} n_c R^{bc} = \alpha ( R^{at} + \beta^a R^{tt} ) \nonumber \\
& = \frac{4}{3} E W^2 v^a + W \Fs^a + \Fs W v^a,
\end{align}
where we have used $\gamma^a_{~b} n_c h^{bc} = - W^2 v^a$.
Finally, the radiation stress tensor as observed by a normal observer is given by a purely spatial projection of (\ref{rad_se}),
\begin{align} \label{Srad}
\Srad^{ab} & \equiv \gamma^a_{~c} \gamma^b_{~d} R^{cd} =  R^{ab} - \alpha n^a R^{bt} - \alpha n^b R^{at} + \alpha^2 n^a n^b R^{tt}  \nonumber \\
& = \frac{4}{3} E W^2 v^a v^b + \frac{1}{3} E \gamma^{ab} +  W \Fs^a v^b + W v^a \Fs^b.
\end{align}
In the above expressions we introduced the bars in order to distinguish these radiation quantities from similar quantities often defined in the 3+1 decomposition of the matter stress-energy tensor.

\subsection{Dynamical equations for the radiation fields}
\label{sec:eqs_dynamics}

We now project the dynamical equations (\ref{eom}) both along the normal and the into spatial slice.  The former will give rise to the radiation energy equation (\ref{tau_dot}) below, while the latter results in the radiation momentum (or flux) equations (\ref{S_dot}).

\subsubsection{The energy equation}
\label{sec:eqs_dynamics_energy}

We start with a normal projection of (\ref{eom}), 
\begin{equation} \label{eom_normal}
n_a \nabla_b R^{ab} = \nabla_b (n_a R^{ab}) - R^{ab} \nabla_b n_a = - n_a \Gst^a = \Gs.
\end{equation}
Applying the identity
\begin{equation}
\nabla_a V^a = \frac{1}{\sqrt{|g|}} \partial_a \left( \sqrt{|g|} \, V^a \right)
\end{equation}
twice -- once for the covariant derivative $\nabla_a$ associated with the spacetime metric $g_{ab}$ and its determinant $g$, and once for the covariant derivative $\Dref_i$ associated with the reference metric $\hat \gamma_{ij}$ and its determinant $\hat \gamma$ -- we may rewrite the first term in the first equality of (\ref{eom_normal}) as
\begin{align} \label{eq1}
&\nabla_b (n_a R^{ab}) = \frac{1}{\sqrt{-g}} \partial_b \left( \sqrt{-g} n_a R^{ab} \right) \\
 & =  \frac{1}{\sqrt{-g}}   \left\{ \partial_t \left( \sqrt{-g} n_a R^{at} \right) +   \partial_i \left( \sqrt{-g} n_a R^{ai} \right) \right\} \nonumber \\
 &=  - \frac{1}{\alpha \sqrt{\gamma}} \left\{ \partial_t \left( \sqrt{\gamma} \,\alpha^2 R^{tt} \right) +  \partial_i \left( \sqrt{\gamma} \,\alpha^2 R^{it} \right)  \right\} \nonumber \\
&= - \frac{1}{\alpha \sqrt{\gamma / \hat \gamma}} \left\{ \partial_t \left( \sqrt{\gamma / \hat \gamma} \, \alpha^2 R^{tt} \right) +  \Dref_i \left( \sqrt{\gamma / \hat \gamma} \, \alpha^2 R^{it} \right) \right\}.   \nonumber
\end{align}
Here we have used $g = -\alpha \gamma$, where $\gamma$ is the determinant of the spatial metric $\gamma_{ij}$.    We have also assumed that the determinant of the reference metric, $\hat \gamma$, is independent of time, which would be easy to generalize if desired.   Inserting (\ref{eq1}) into (\ref{eom_normal}) we obtain
\begin{equation} \label{tau_dot}
\fbox{$
\partial_t \bar \tau + \Dref_i f_{\bar \tau}^i = s_{\bar \tau} - \alpha \sqrt{\gamma / \hat \gamma}\, \Gs,
$}
\end{equation}
where we have defined the radiation energy density variable
\begin{equation} \label{tau}
\bar \tau \equiv \sqrt{\gamma / \hat \gamma}\, \alpha^2 R^{tt} = \sqrt{\gamma / \hat \gamma} \, \rhorad,
\end{equation}
its associated energy  flux,
\begin{align} \label{f_tau}
f_{\bar \tau}^i & \equiv \sqrt{\gamma / \hat \gamma} \, \alpha^2 R^{it} = \sqrt{\gamma / \hat \gamma} \, (\alpha \jrad^i - \rhorad \beta^i ) \\
& = \bar \tau (\alpha v^i - \beta^i) + \alpha \sqrt{\gamma / \hat \gamma} \,\left( \frac{1}{3} E v^i - W \Fs v^i + W \Fs^i \right) \nonumber 
\end{align}
as well as the source term
\begin{align} \label{s_tau}
s_{\bar \tau} & \equiv - \alpha \sqrt{\gamma / \hat \gamma}\, R^{ab} \nabla_b n_a 
= \alpha \sqrt{\gamma / \hat \gamma}\, R^{ab} ( K_{ba} + n_b a_a) \nonumber \\ 
& = \sqrt{\gamma / \hat \gamma}\, ( \alpha \Srad^{ij} K_{ij} - \jrad^i \partial_i \alpha ).
\end{align}
In the last equation 
\begin{equation}
K_{ab} \equiv - \gamma_a^{~c} \gamma_b^{~d} \nabla_c n_d = - \nabla_a n_b - n_a a_b
\end{equation}
is the extrinsic curvature, and 
\begin{equation}
a_b \equiv n^a \nabla_a n_b = \gamma_b^{~c} \partial_c \ln \alpha
\end{equation} 
the acceleration of the normal observer.

We note that, in the reference-metric formalism, all quantities are defined using ratios between determinants, rather than just the determinants themselves, and are therefore tensor-densities of weight zero.  We will discuss some other computational advantages of the reference-metric formalism in Section \ref{sec:numerics_OS_early_analytical}  below.

\subsubsection{The momentum equation}
\label{sec:eqs_dynamics_momentum}

We now take a spatial projection of (\ref{eom}), which yields
\begin{equation} \label{eom_spatial}
\gamma_{ia} \nabla_b R^{ab} = - \gamma_{ia} \Gst^a = - \Gs_i.
\end{equation}
We first rewrite
\begin{equation}
\gamma_{ia} \nabla_b R^{ab} = g_{ia} \nabla_b R^{ab} = \nabla_b (R_i^{~b})
\end{equation}
and then use the identity
\begin{equation}
\nabla_b T_a^{~b} = \frac{1}{\sqrt{|g|}} \partial_b \left( \sqrt{|g|} \, T_a^{~b} \right) - T_c^{~b} \, \4G^c_{ab}
\end{equation}
twice to find
\begin{align} \label{eq3}
& \gamma_{ia} \nabla_b R^{ab} \nonumber \\
& ~~~   =  \frac{1}{\alpha \sqrt{\gamma / \hat \gamma}} 
\left\{ \partial_t \left(\alpha \sqrt{\gamma / \hat \gamma} \, R_i^{~t} \right) + \Dref_j \left(\alpha \sqrt{\gamma / \hat \gamma} \, R_i^{~j} \right) \right\} \nonumber \\
& ~~~~~~ + R_j^{~k} \hat \Gamma^j_{ki} - R_c^{~b} \, \4G^c_{bi}, 
\end{align}
where the $\4G^c_{ab}$ are the Christoffel symbols associated with the spacetime metric $g_{ab}$, and 
the $\hat \Gamma^j_{ki}$ are those associated with the reference metric $\hat \gamma_{ij}$. Inserting (\ref{eq3}) into (\ref{eom_spatial}) we obtain
\begin{equation} \label{S_dot}
\fbox{$
\partial_t \bar S_i + \Dref_j (f_{\bar S})_i^{~j} = (s_{\bar S})_i  - \alpha \sqrt{\gamma / \hat \gamma} \, \Gs_i,
$}
\end{equation}
where we have defined the radiation momentum density, or flux, variable
\begin{equation} \label{S}
\bar S_i \equiv \alpha \sqrt{\gamma / \hat \gamma} \, R_i^{~t} = \sqrt{\gamma / \hat \gamma} \, \jrad_i,
\end{equation}
its associated momentum flux
\begin{align} \label{S_flux}
(f_{\bar S})_i^{~j} & \equiv \alpha \sqrt{\gamma / \hat \gamma} \, R_i^{~j} = \sqrt{\gamma / \hat \gamma} \, ( \alpha \Srad_i^{~j} - \jrad_i \beta^j) \nonumber \\
& = \bar S_i ( \alpha v^j - \beta^j) \\ 
&~~~~~~+ \alpha \sqrt{\gamma / \hat \gamma} \,\left(\frac{1}{3} E \delta_i^{~j} - W \Fs v_i v^j + W v_i \Fs^j \right), \nonumber
\end{align}
and the source term
\begin{equation} \label{s_S}
(s_{\bar S})_i \equiv \alpha \sqrt{\gamma / \hat \gamma} \, ( R_c^{~b} \4G^c_{bi} - R_j^{~k} \hat \Gamma^j_{ki} ).
\end{equation}
We now write
\begin{equation}
R_c^{~b} \4G^c_{bi} - R_j^{~k} \hat \Gamma^j_{ki} = R^{cb} \4G_{cbi} - R^{ck} g_{jc} \hat \Gamma^j_{ki},
\end{equation}
expand $R^{ab}$ into its projections (\ref{rhorad}), (\ref{jrad}) and (\ref{Srad}), and use
\begin{equation}
\4G_{(bc)i} = \partial_i g_{bc} = - g_{db} g_{ec} \partial_i g^{de}
\end{equation}
(where $()$ denotes symmetrization) to rewrite the source term (\ref{s_S}) as
\begin{equation}
(s_{\bar S})_i = \sqrt{\gamma / \hat \gamma} \left( - \rhorad \, \Dref_i \alpha + \jrad_j \Dref_i \beta^j + \frac{1}{2} \alpha \bar S^{jk} \Dref_i \gamma_{jk} \right) 
\end{equation}
(see Section III.B in MBM; also note that $\jrad_t = \jrad_i \beta^i$).  

For most numerical applications, a natural choice for the reference metric $\hat \gamma_{ij}$ is the flat (spatial) metric in whatever coordinate system used.  If so, \eqs~(\ref{tau_dot}) and (\ref{S_dot}) reduce to familiar expressions (e.g.~\eqs~(35) and (38) of FLLS) when evaluated in Cartesian coordinates, for which $\hat \gamma = 1$ and $\Dref_i = \partial_i$. In curvilinear coordinates, we evaluate the flux terms in both equations by writing, for example, $\Dref_i f_{\bar \tau}^i = \partial_i f_{\bar \tau}^i + f_{\bar \tau}^j \hat \Gamma^i_{ji}$, where the Christoffel symbols  $\hat \Gamma^i_{jk}$ are known analytically.  We then move these Christoffel terms to the right-hand sides of the equations, as discussed in MBM, so that they act as source terms.

\eqs~(\ref{tau_dot}) and (\ref{S_dot}) now form the dynamical equations for the conserved radiation variables $\bar \tau$ and $\bar S_i$; in a numerical simulation they have to be solved together with the equations for the gravitational fields, relativistic hydrodynamics and any other fields or sources that are being considered.  We present a simple analytical example, also highlighting some advantages of the reference-metric formulation, in Section \ref{sec:numerics_OS_early_analytical}.

%
%

\subsection{Recovery}
\label{sec:eqs_recovery}


Solving \eqs~(\ref{tau_dot}) and (\ref{S_dot}) yields the conserved radiation variables $\bar \tau$ and $\bar S_i$.  In the course of the dynamical evolution, however, we also need the primitive variables $E$ and $F^a$ -- or, equivalently, $E$, $\Fs$ and $\Fs^i$.  The latter variables therefore have to be recovered from the conserved variables.  For the hydrodynamical variables, a similar recovery step generally requires a numerical iteration, but for the radiation equations treated here the recovery can be accomplished algebraically, as was the case in FLLS.

We start by using (\ref{rho_bar}) in (\ref{tau}),
\begin{equation} \label{recov_1}
\bar \tau = \sqrt{\gamma / \hat \gamma} \left( \frac{1}{3} ( 4 W^2 - 1) E + 2 W \Fs \right). 
\end{equation}
Next we compute the contraction $n_a u_b R^{ab}$ twice; once expressing $R^{ab}$ as in (\ref{rad_se}), i.e.~projected with respect to $u^a$,
\begin{equation}
n^a u^b R_{ab} = W E + \Fs,
\end{equation}
and once expressing $R^{ab}$ in terms of its spatial projections,
\begin{align}
n^a u^b R_{ab} & = n^a g_c^{~b} u^c R_{ab} = n^a (\gamma_c^{~b} - n_c n^b) u^c R_{ab} \nonumber \\
& = \rhorad W - \jrad_c u^c = W (\rhorad - \jrad_i v^i).
\end{align}
Multiplying both expressions with $\sqrt{\gamma / \hat \gamma}$ and equating them yields
\begin{equation} \label{recov_2}
W (\bar \tau - \bar S_i v^i) = \sqrt{\gamma / \hat \gamma} \, ( W E + \Fs).
\end{equation}
This is equivalent to Eq.~(66) in FLLS, once the different definitions of the spatial velocity $v^i$ have been taken into account (see Eq.~\ref{v_FLLS} above).

Eqs.~(\ref{recov_1}) and (\ref{recov_2}) now provide two equations for two unknowns $E$ and $\Fs$ that can be solved directly given values of the conserved variables $\bar \tau$ and $\bar S_i$.

Finally, we insert (\ref{jrad}) into (\ref{S}),
\begin{equation}
\bar S^i = \sqrt{\gamma / \bar \gamma} \left( \frac{4}{3} E W^2 v^a + W \Fs^a + W \Fs v^a \right),
\end{equation}
and solve for $\Fs^a$ to obtain
\begin{equation} \label{f_recovery}
\Fs^a = \frac{1}{W \sqrt{\gamma / \hat \gamma}} \bar S^i - \frac{4}{3} E W v^a - \Fs v^a
\end{equation}
(compare Eq.~67 in FLLS).  This completes the recovery of the primitive variables $E$, $\Fs$, and $\Fs^a$ from the conserved variables $\bar \tau$ and $\bar S_i$.  

While the recovery of the primitive radiation variables involves algebraic equations only, the solution may nevertheless be affected by significant numerical error, especially at large optical depths.  This is because, in such regions, the flux variables $\Fs$ and $\Fs^i$ will often be much smaller than the radiation energy density $E$ as well as the conserved quantities $\bar \tau$ and $\bar S_i$.  In this case, the flux variables are computed as the small differences between (potentially) much larger numbers, which generally leads to increased numerical error.  We will discuss a concrete example in Section \ref{sec:numerics_OS_early_numerical} below.

\section{Numerical examples}
\label{sec:numerics}

\subsection{Numerical implementation}
\label{sec:numerics_implementation}

Most features of our numerical implementation have been described in \cite{BauMCM13,MonBM14,BauMM15}.  Specifically, we use a reference-metric approach \cite{BonGGN04,ShiUF04,Bro09,Gou12} to express the Baumgarte-Shapiro-Shibata-Nakamura (BSSN) formulation \cite{NakOK87,ShiN95,BauS98} of Einstein's equations as well as the equations of relativistic hydrodynamics in spherical polar coordinates.  Specifically, we adopt the flat metric in spherical polar coordinates as our reference metric $\hat \gamma_{ij}$.  We rescale all tensorial quantities with appropriate powers of $r$ and $\sin \theta$ so that all singular terms can be treated analytically.  For example, for a vector with covariant components we write
\begin{equation} \label{rescaling_vector}
\bar S_i = \left(
\begin{array}{c}
\tilde S_r \\ r \, \tilde S_\theta \\ r \sin \theta \, \tilde S_\varphi
\end{array}
\right)
\end{equation}
and evolve the variables $\tilde S_i$ in our code.  For vectors with contravariant components we divide by similar factors; for the fluxes  $(f_{\bar S})_i^{~j}$ in (\ref{S_flux}) with mixed indices we write
\begin{equation} \label{rescaling_tensor}
(f_{\bar S})_i^{~j} = \left(
\begin{array}{ccc}
(\tilde f_{\bar S})_r^{~r} & (\tilde f_{\bar S})_r^{~\theta}/r & (\tilde f_{\bar S})_r^{~\varphi} / (r \sin \theta) \\
r \, (\tilde f_{\bar S})_\theta^{~r} & (\tilde f_{\bar S})_\theta^{~\theta} & (\tilde f_{\bar S})_\theta^{~\varphi} / \sin \theta \\
r \sin \theta \, (\tilde f_{\bar S})_\varphi^{~r} & \sin \theta \, (\tilde f_{\bar S})_\varphi^{~\theta} & (\tilde f_{\bar S})_\varphi^{~\varphi} \\
\end{array}
\right).
\end{equation}
We impose parity boundary conditions to allow finite-differencing across the origin and the axis (see, e.g., Table I in \cite{BauMCM13}), and Robin-type conditions on the outer boundaries.

The latest version of our code uses fourth-order differencing for all spatial derivatives in Einstein's field equations, together with a fourth-order Runge-Kutta time integrator \cite{BauGH19}.  We solve the equations of relativistic hydrodynamics using an HLLE approximate Riemann solver \cite{HarLL83,Ein88}, together with a simple monotonized central-difference limiter reconstruction scheme \cite{Van77}.  The latter is second-order accurate in most regions, but reduces to first order close to discontinuities or extrema.  More accurate schemes are used by many groups (e.g.~\cite{ShiF05,YamST08,Jesetal20}; see also \cite{Tor99,RezZ13}), but are not needed for the numerical examples presented here.

We have now implemented the equations of radiation hydrodynamics, in the gray, optically-thick two-moment approximation of Section \ref{sec:eqs} above, in the exact same computational framework as those of relativistic hydrodynamics, allowing for fully relativistic radiation hydrodynamics simulations in spherical polar coordinates.  As numerical demonstrations we consider flat spacetime tests in Sect.~\ref{sec:numerics_flat}, and heated Oppenheimer-Snyder collapse in Sect.~\ref{sec:numerics_OS}.

\subsection{Flat spacetime tests}
\label{sec:numerics_flat}

\begin{table*}
\begin{tabular}{c|c|c|c||c|c|c|c||c|c|c|c}
	&  & & & \multicolumn{4}{c||}{left state} & \multicolumn{4}{c}{right state} \\
\hline
type 	& $\kappaa$ & $\Gamma$ & $a_R m^4$ & $\rho_0$ & $P$ & $u^z$ & $E$ &  $\rho_0$ & $P$ & $u^z$ & $E$ \\
\hline
\hline
continuous
 & 0.08 &  5/3 &  $1.39 \times 10^{8}$  & 1.0 & $6 \times 10^{-3}$     & 0.69                       & $0.18$ 
                    & 3.65 & $3.59 \times 10^{-2}$  & $0.189$ & $1.30$ \\  

shock
 & 0.24 &  5/3 &  $1.24 \times 10^{10}$ & 1.0 & $3 \times 10^{-5}$     & 0.015                       & $1.0 \times 10^{-8}$ 
                    & 2.4 & $1.61 \times 10^{-4}$  & $6.25 \times 10^{-3}$ & $2.51 \times 10^{-7}$ \\  
\end{tabular}
\caption{Left and right states for the flat spacetime tests of Sect.~\ref{sec:numerics_flat}.  The continuous solution in the top row corresponds to Test 4 in Table I of FLLS, while the shock solution in the bottom row corresponds to their Test 1 (except that we use $\kappaa = 0.24$).}
\label{Table:1}
\end{table*}

Stationary and slab-symmetric solutions to the equations of relativistic radiation hydrodynamics in flat (Minkowski) spacetimes can be derived by assuming that the solutions to \eqs~(\ref{eom}), 
as well as the equations of relativistic hydrodynamics, are independent of time, and depend on one spatial Cartesian coordinate only, say $z$ (see \cite{ZelR66,MihWM84}).  Further assuming a $\Gamma$-law equation of state, 
\begin{equation} \label{gamma_law}
P = (\Gamma - 1) \epsilon \rho_0,  
\end{equation}
the equations reduce to a set of five coupled ordinary differential equations that can be solved as discussed in Appendix C of FLLS \footnote{We would like to thank Yuk Tung Liu for pointing out that the first term in the second line of \eq~(C20) in FLLS misses an overall factor $u^0_R$.}.  We will assume that $\kappas = 0$, and that $\kappaa$ is constant.  We then transform these semi-analytical solutions to spherical polar coordinates, and adopt the resulting data as initial data for our dynamical evolutions.  Given that the data describe stationary solutions, any departure from the initial data serves as a measure of numerical error.

\subsubsection{Continuous solutions}
\label{sec:numerics_flat_cont}

Continuous semi-analytic solutions can be constructed by adopting boundary conditions at the lower boundary for $E$ and $F^z \ll E$, as well as for the fluid's rest-mass density $\rho_0$, pressure $P$, and four-velocity $u^z$, and integrating to larger values of $z$.  In particular, we assume that the radiation is in thermal equilibrium with the fluid at the lower boundary, i.e.~$E = 4 \pi B = a_R T^4 = a_R m^4 (P / \rho_0)^4$.  Here we have adopted the Maxwell-Boltzmann ideal gas law $P = \rho_0 T / m$, where $m$ is the mean mass of the fluid particles, and where we have chosen units in which Boltzmann's constant is unity, $k_B = 1$.   Since no shocks are encountered in this test, we replaced the monotonized central-difference limiter reconstruction scheme with simple quadratic interpolation, and therefore expect second-order convergence for these simulations.

\begin{figure}[t]
\includegraphics[width = 0.4 \textwidth]{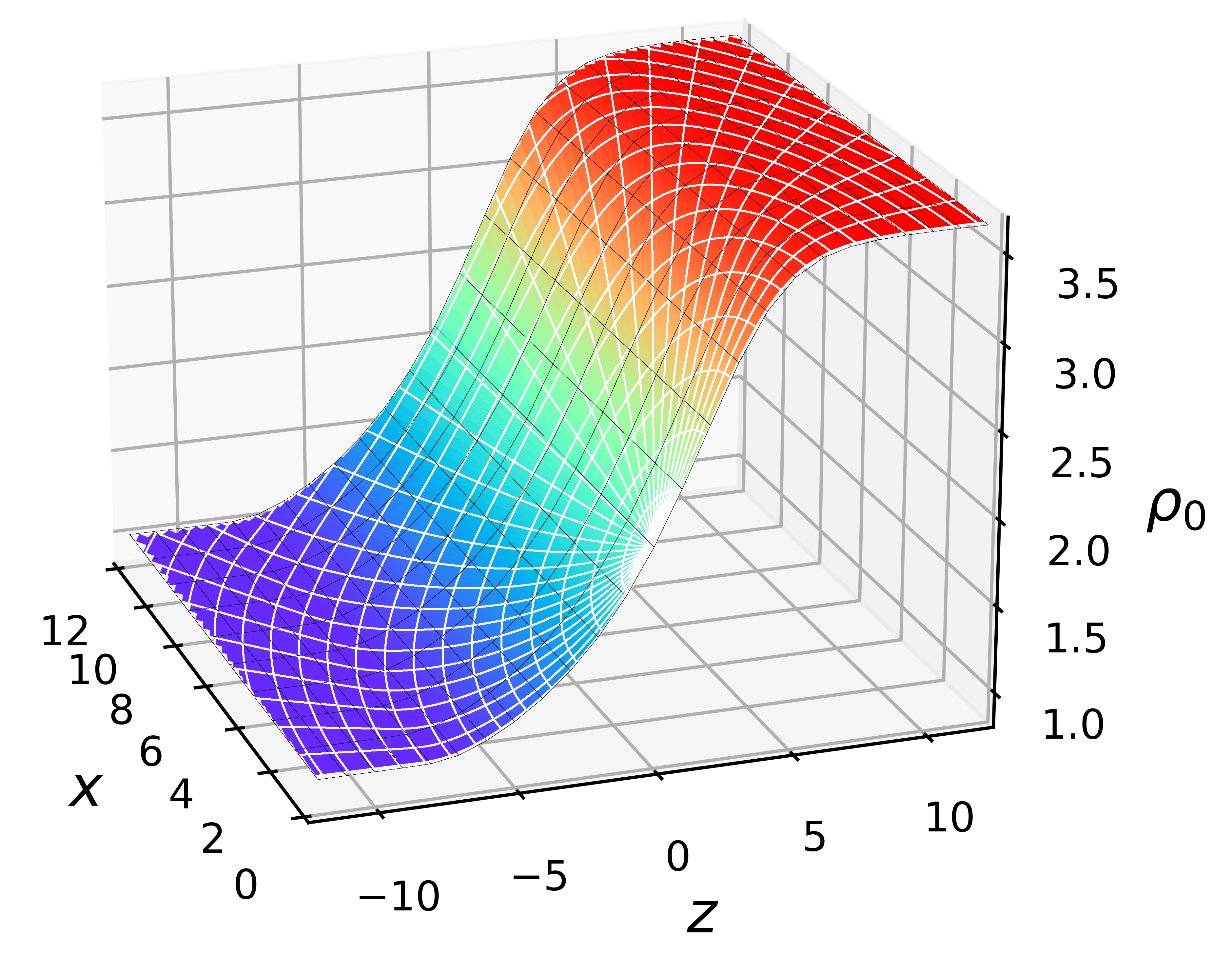}
\includegraphics[width = 0.4 \textwidth]{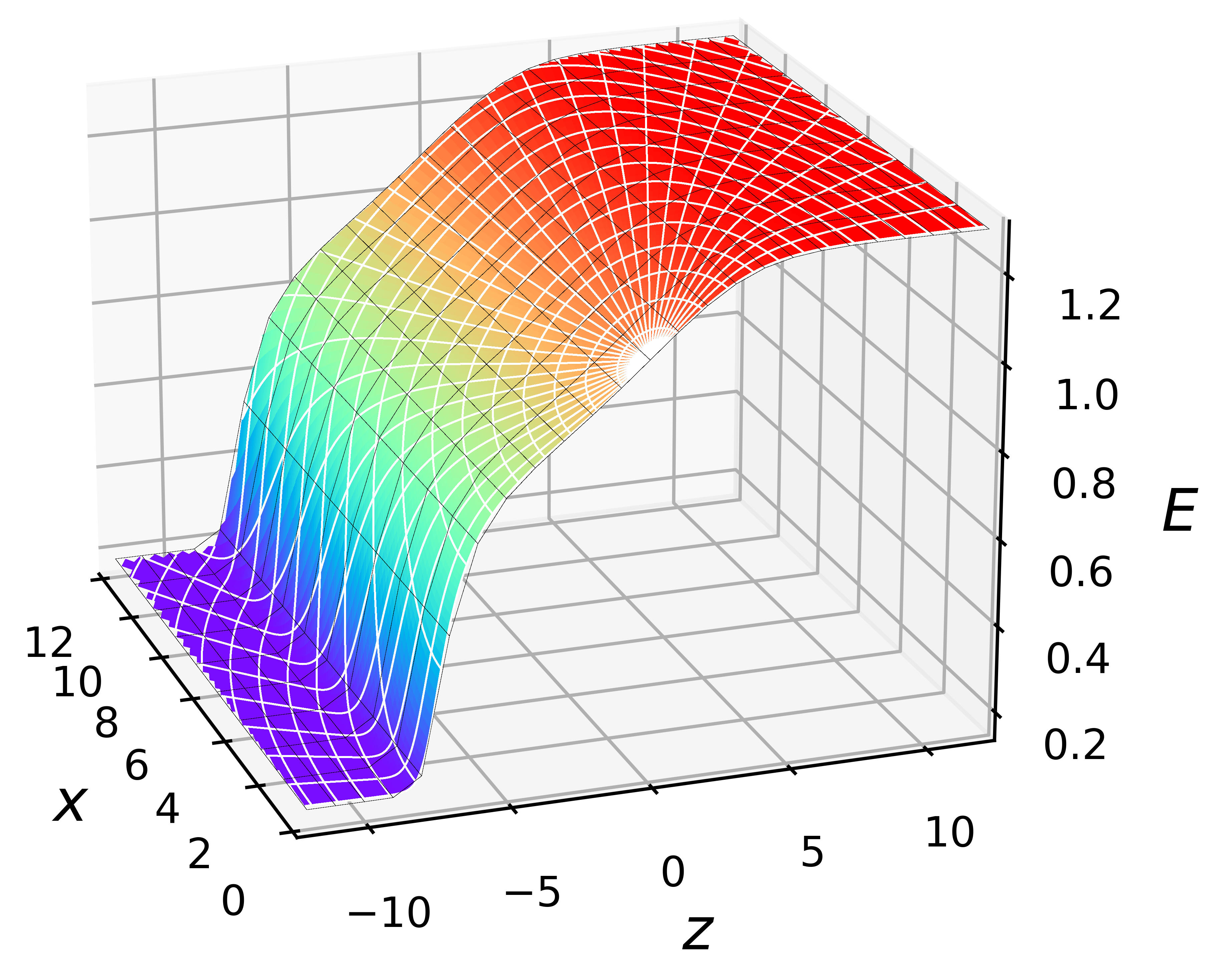}
\caption{A continuous flat-spacetime solution, showing the fluid rest-mass density $\rho_0$ in the top panel and the radiation energy density $E$ in the bottom panel.  The black rectangular grid represents the semi-analytical solution for the data in the top row of Table \ref{Table:1}, while the colored surface shows the numerical solution at time $t = 10.053$, obtained with $N_r = 320$ radial and $N_\theta = 120$ angular grid points, with the grid extending to $r_{\rm out} = 24$.  The white lines represent our spherical polar coordinate system, showing every 12-th radial and every 4-th angular grid line.}
\label{Fig:1}
\end{figure}

As an example, we show results for the boundary values listed in the top row of Table \ref{Table:1}, which correspond to Test 4 listed in Table I of FLLS.  We show profiles of this solution in Fig.~\ref{Fig:1}, comparing the numerical solution at time $t = 10.053$ (displayed as the colored surface with spherical polar coordinate lines) with the semi-analytical solution (represented by rectangular grid).  It is difficult to see any difference in these plots.

\begin{figure}[t]
\includegraphics[width = 0.45 \textwidth]{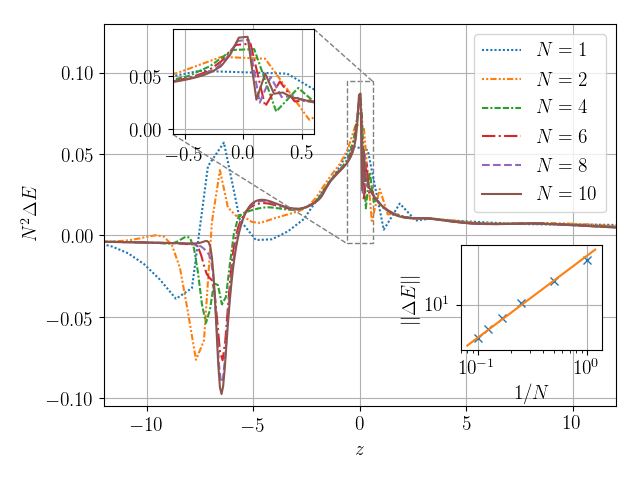}
\caption{Convergence test for the continuous solution shown in Fig.~\ref{Fig:1}, except that we have also boosted the solution with a speed $\beta = 0.1$ in the $z$-direction for this test.  The different lines show differences $N^2 \Delta E$, rescaled assuming second-order convergence, between the numerical and semi-analytical solutions.  The different lines show interpolations to the $z$-axis, for grids with $N_r = 32 N$ radial and $N_\theta = 12 N$ grid points, and with the numerical grid extending to $r_{\rm out} = 24$.   The top left inset shows the behavior in the vicinity of the center, demonstrating second-order convergence even in the presence of the coordinate singularities at the origin.  The bottom right inset shows results for the norm $||\Delta E|| \equiv \int | \Delta E | dV$, integrated to a radius of $r = 12$.  The solid line in this inset represents a power-law $(1 / N)^2$.}
\label{Fig:2}
\end{figure}

In Fig.~\ref{Fig:2} we show a convergence test for this setup, except that we have also boosted the solution with a speed $\beta = 0.1$ in the positive $z$-direction for this test.  We interpolate our numerical solutions to the $z$-axis, then compute the difference $\Delta E$ between this numerical solution and the semi-analytical solution, and finally multiply these differences with $N^2$.  The plot shows that these rescaled differences $N^2 \Delta E$ converge, establishing the expected second-order convergence.  The bottom right inset shows that integrals of the numerical error also decrease with $N^{-2}$, as expected.

\subsubsection{Discontinuous solutions}
\label{sec:numerics_flat_discont}

Solutions featuring a shock discontinuity, on the other hand, can be constructed by assuming that the radiation is in thermal equilibrium with the fluid at both the lower and the upper boundary.  As discussed in Appendix C of FLLS, a ``shooting method" can then be employed to integrate the equations from both boundaries to the location of a shock discontinuity at $z = z_{\rm shock}$, and imposing matching conditions there \footnote{It appears difficult to construct these solutions exactly as described in Appendix C of FLLS, because the solutions feature exponential growth approaching the shock discontinuity from both sides.   For large optical depths, the iteration used to match these solutions at the discontinuity requires more precision than can be achieved even with long double precision.   We avoided these problems by restricting our optical depths to values slightly smaller than those reported by FLLS, choosing $\kappaa = 0.24$ rather than their value of $\kappaa = 0.4$.}.      

\begin{figure}[t]
\includegraphics[width = 0.4 \textwidth]{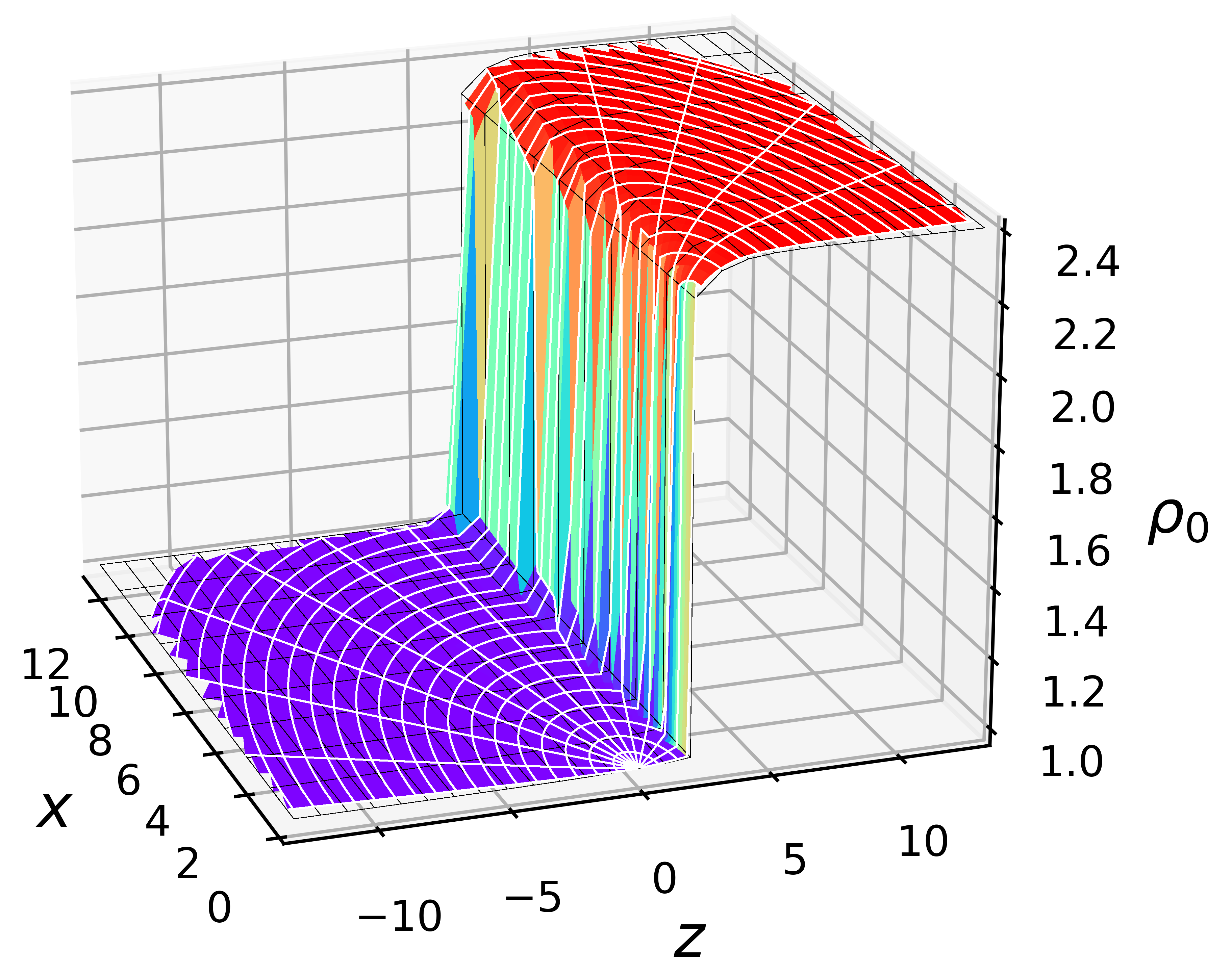}
\includegraphics[width = 0.4 \textwidth]{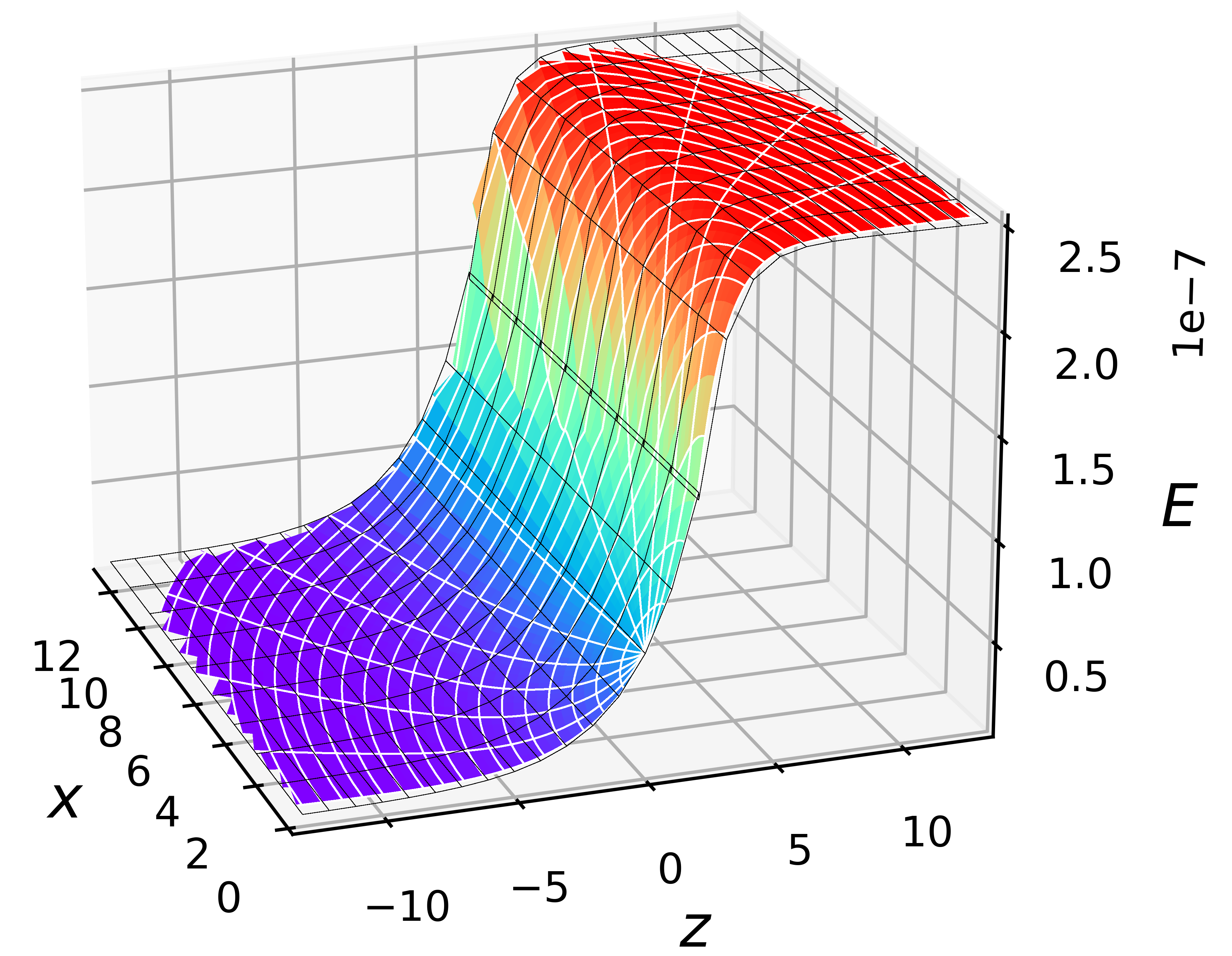}
\caption{Same as Fig.~\ref{Fig:1}, except for a solution featuring a shock discontinuity (see the bottom row in Table \ref{Table:1}).   For this test, performed with $N_r = 256$ radial and $N_\theta = 48$ angular gridpoints, and the outer boundary at $r_{\rm out} = 16$, we placed the shock discontinuity at $z  = 2$ rather than at $z = 0$, so that the shock front does not coincide with the symmetry plane of the coordinate system.}
\label{Fig:3}
\end{figure}

As an example, we show profiles of the fluid rest-mass density $\rho_0$ and the radiation energy density $E$ at $t = 10.053$ in Fig.~\ref{Fig:3}, demonstrating that the shock discontinuity is well-resolved, even when the shock front does not coincide with a coordinate plane.

\subsection{Heated Oppenheimer-Snyder collapse}
\label{sec:numerics_OS}

An analytical solution describing ``heated Oppenheimer-Snyder collapse", i.e.~collapse of a homogeneous dust sphere to a black hole (see \cite{OppS39}) with radiation, has been derived in \cite{Sha89} (see also \cite{Sha96} and \cite{BauS10} for a summary).  This solution makes several assumptions that are realized only approximately in numerical simulations that adopt a two-moment radiation formalism.  One of these assumptions is that all pressure and radiation terms are sufficiently small so that they do not affect the spacetime and dust evolution; another assumption is that the radiative processes can be described in the relativistic diffusion limit (see Appendix A.1 in FLLS).  The former condition can be met in a numerical simulation by making suitable choices for the equation of state and the initial data.  While the latter approximation is quite adequate during most of the evolution for a star of sufficiently large optical depth, it is violated during an initial transient phase, lasting a few light-travel times across the initial dust sphere, after which it improves in accuracy.  In this Section we carefully discuss the resulting transition from the initial data to post-transient diffusion solution.  An exact numerical solution has been obtained by integrating the Boltzmann equation without approximation in \cite{Sha96}.

\subsubsection{Oppenheimer-Snyder collapse}
\label{sec:numerics_OS_OS}

\begin{figure}[t]
\includegraphics[width = 0.45 \textwidth]{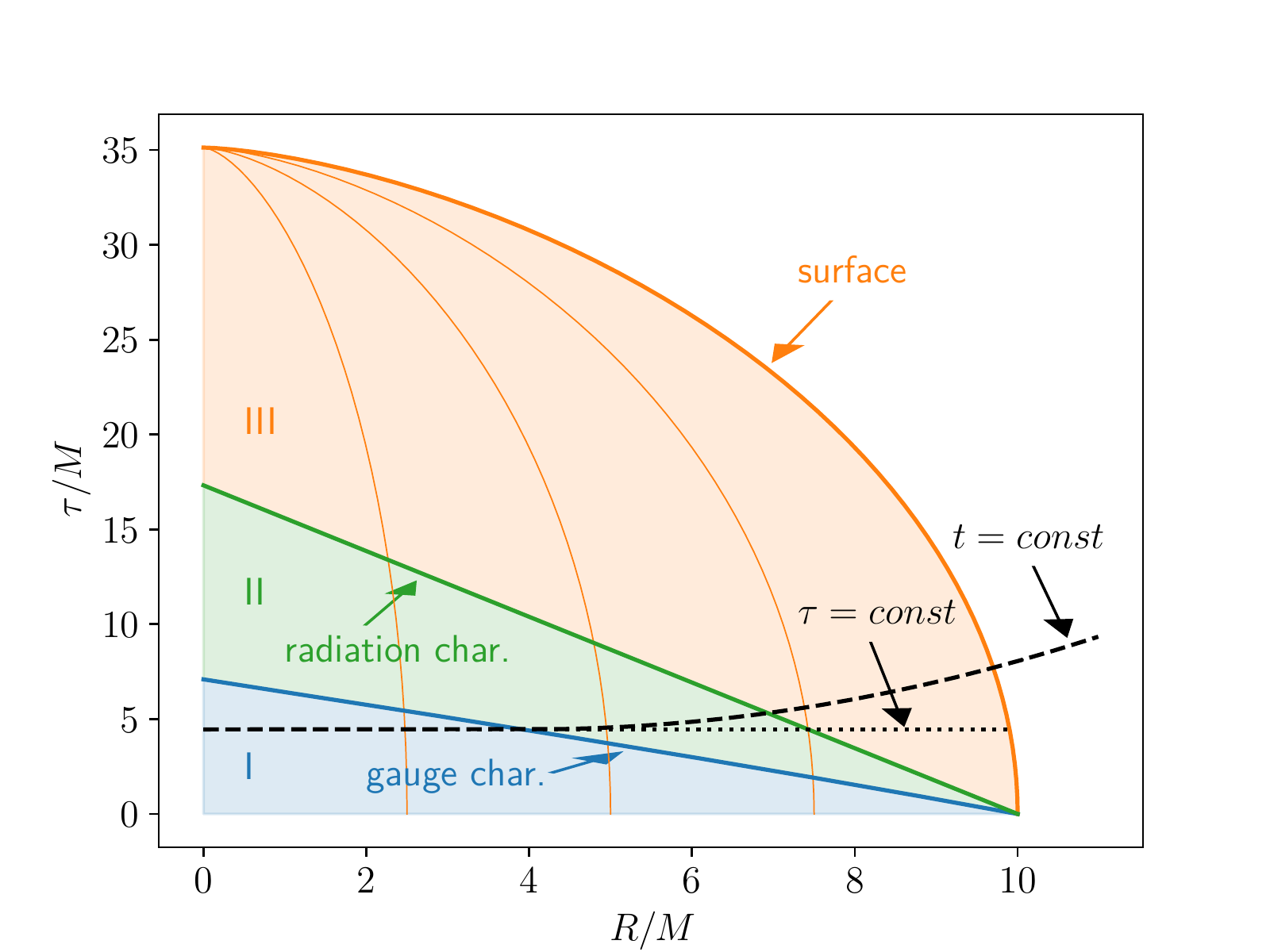}
\caption{A schematic spacetime diagram for Oppenheimer-Snyder collapse.  The (red) lines starting out vertically at $\tau = 0$ trace the worklines of selected dust particles, with the thick line marking the surface.  The (black) dotted horizontal line shows a surface of constant proper time $\tau$ (where $\tau$ is measured by observers comoving with the dust), while the (black) dashed line sketches a surface of constant coordinate time $t$.  Also included are two characteristics originating at the stellar surface and traveling inwards; one for the radiation field, and one for gauge perturbations (see text for details). }
\label{Fig:OS}
\end{figure}

Oppenheimer-Snyder collapse describes the collapse from rest of a constant-density dust sphere to a black hole \cite{OppS39}.  An analytical solution for this collapse can be constructed by matching a closed-Friedmann solution for the stellar interior to a Schwarzschild solution for the exterior.  Expressed in Gaussian normal coordinates, the interior line element is given by
\begin{equation} \label{OS_g}
ds^2 = - d\tau^2 + a^2(\tau) ( d \chi^2 + \sin^2 \chi d\Omega^2),
\end{equation}
where $0 \leq \chi \leq \chi_0$ is a radial coordinate that is comoving with the dust particles, $a(\tau)$ a scale factor, and $\tau$ the proper time as observed by observers comoving with the dust.  It is also useful to define the conformal time $\eta$ according to $d\tau = a(\tau) \, d\eta$, in terms of which the scale factor $a$ can be expressed as
\begin{equation}
a = \frac{1}{2} a_m \left(1 + \cos \eta \right),
\end{equation}
and the proper time $\tau$ as 
\begin{equation}
\tau = \frac{1}{2} a_m \left(\eta + \sin \eta \right)
\end{equation}
with $0 \leq \eta \leq \pi$.  Matching this interior solution to a Schwarzschild exterior solution at the stellar surface results in relations for the initial scale factor $a_m = a(0)$,
\begin{equation}
a_m = \left( \frac{R_0^3}{2 M} \right)^{1/2}
\end{equation}
and the maximum value of the radial coordinate $\chi$,
\begin{equation}
\chi_0 = \sin^{-1} \left( \sqrt{2 M / R_0} \right).
\end{equation}
Here $R_0$ is the initial areal radius of the dust cloud and $M$ its gravitational mass (see also Section 1.4 in \cite{BauS10} as well as \cite{StaBBFS12}).

The dust's rest-mass density $\rho_0 = u_a u_b T^{ab}$, where $u^a$ is the dust particles' four-velocity and $T^{ab}$ the stress-energy tensor, remains homogenous on each slice of constant proper time $\tau$ and is given by
\begin{equation} \label{rho_0_OS}
\frac{\rho_0(\tau)}{\rho_0(0)} = \left( \frac{a_m}{a(\tau)} \right)^3.
\end{equation} 
The initial rest-mass density $\rho_0(0)$ is related to the dust cloud's mass and initial radius by
\begin{equation}
M = \frac{4 \pi}{3} \rho_0(0) R_0^3.
\end{equation}

\subsubsection{Heated Oppenheimer-Snyder collapse: diffusion approximation}
\label{sec:numerics_OS_diff}

Given a suitable number of approximations, the radiation emerging from a ``heated" Oppenheimer-Snyder collapse can be described analytically (see \cite{Sha89}).  Specifically, the solutions assumes that neither the spacetime nor the dust evolution are affected by the radiation field, that the radiation is in local thermal equilibrium everywhere (so that $4 \pi B = E = a_R T^4$), and that certain time derivatives can be neglected in an optically thick gas, so that equations (\ref{tau_dot}) and (\ref{S_dot}) can be combined to form a relativistic diffusion equation (see also Appendix A.1 in FLLS).  Further assuming that the initial radiation energy density is constant, $E(0) = E_0$ and that the initial flux vanishes, the analytical solution is given by equations (3.23) and (3.25) in \cite{Sha89}.   In \cite{Sha96}, this analytical solution was compared with an exact numerical solution of the Boltzmann equation (radiation transport equation) without approximation, including the exact boundary conditions at the surface; the two approaches showed very good agreement for the energy density $E(\tau)$ and the emergent flux, i.e.~the radiation momentum density $F^a$ evaluated on the stellar surface.   However, the comparison focussed on the post-transient behavior, after a few light-crossing times across the star.

\begin{figure}[t]
\includegraphics[width = 0.4 \textwidth]{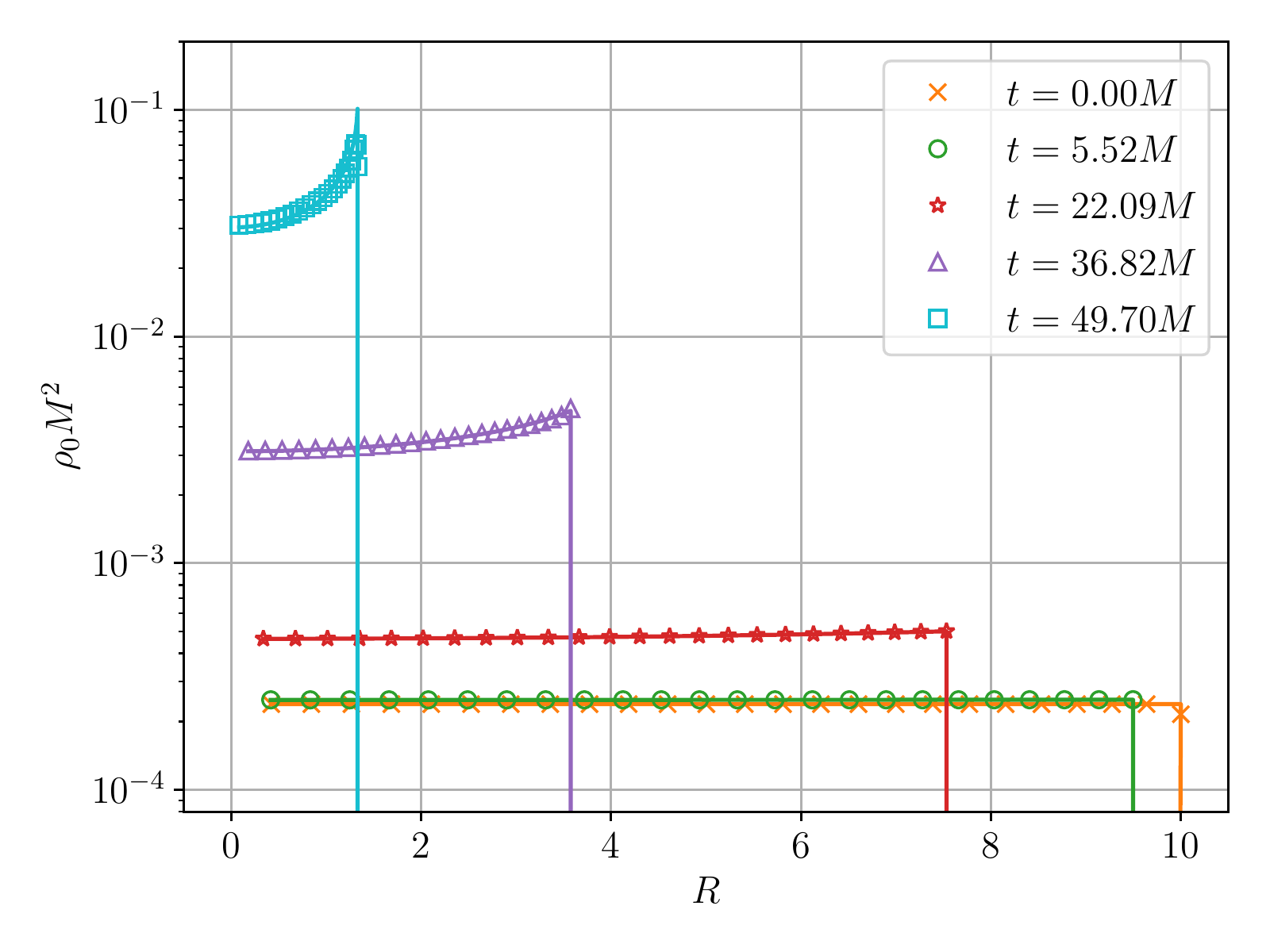}
\includegraphics[width = 0.4 \textwidth]{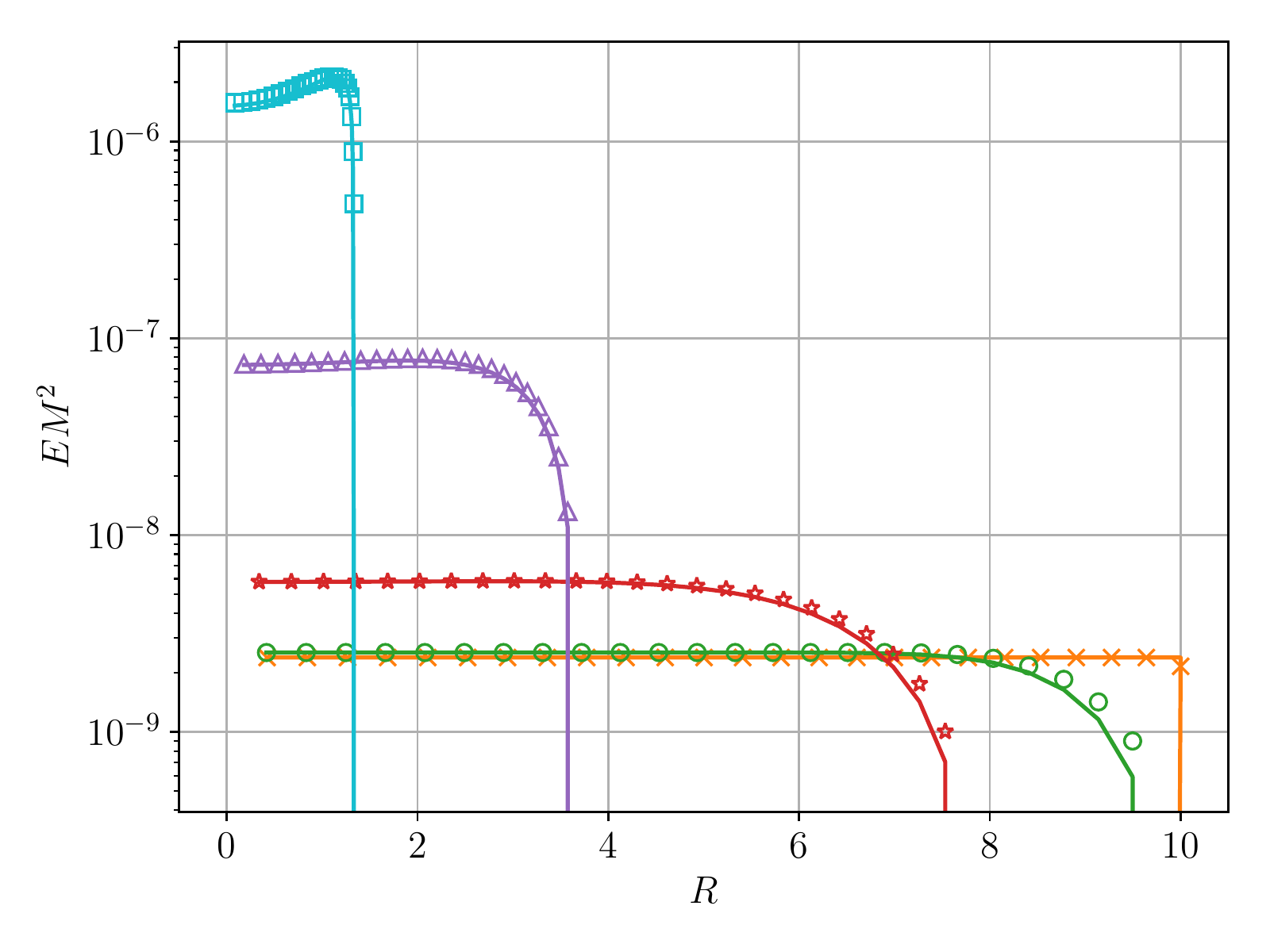}
\includegraphics[width = 0.4 \textwidth]{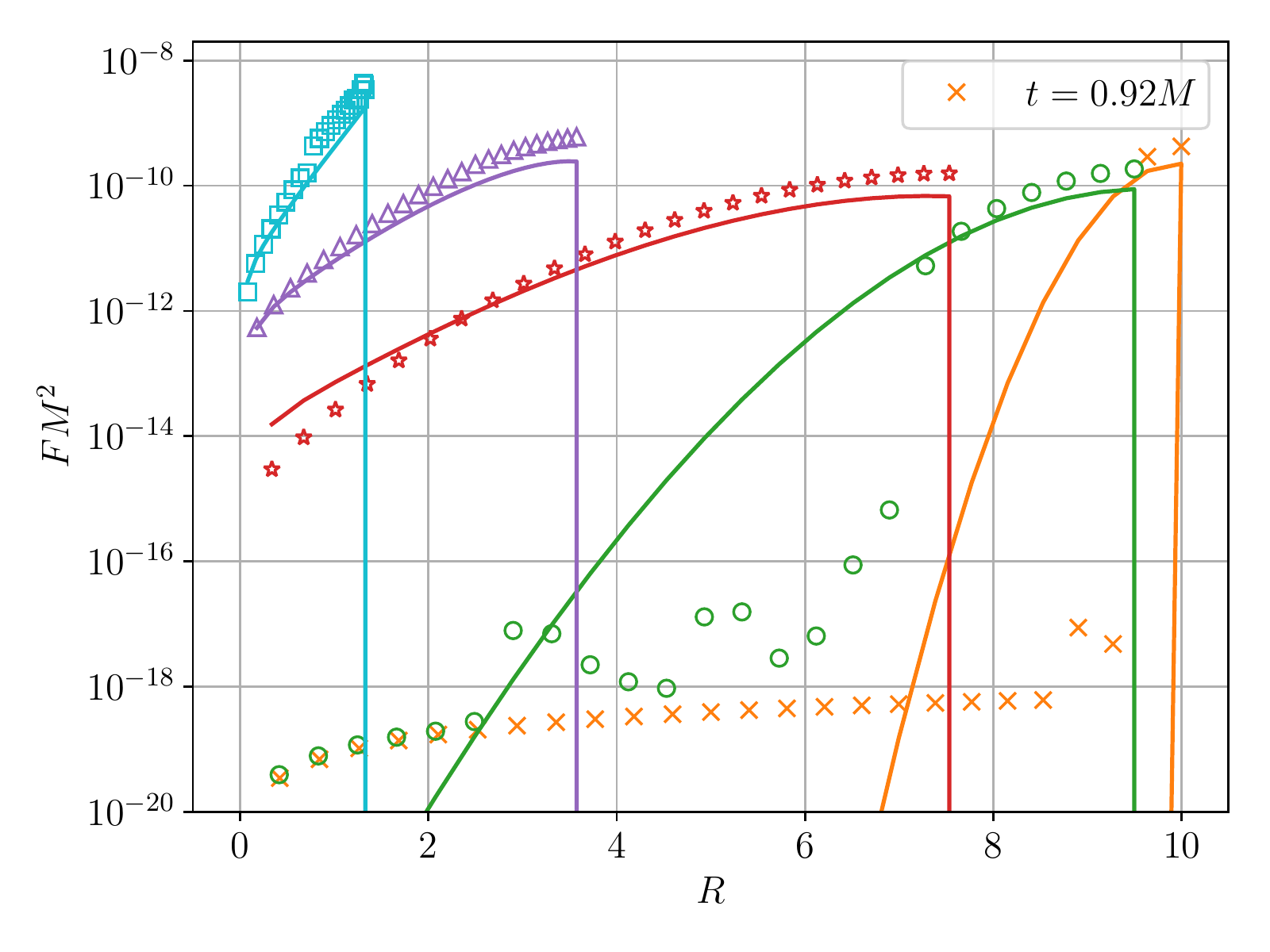}
\caption{A comparison of numerical results and analytical expressions for the rest-mass density $\rho_0$ (top panel), the radiation energy density $E$ (middle panel), and the magnitude of the flux $F \equiv (F_a F^a)^{1/2}$ (bottom panel)  for a heated Oppenheimer with $R_0 = 10 M$ (see text for details).  The markers represent individual Lagrangian fluid tracers (rather than grid points), while the solid lines represent the analytical solution in the diffusion approximation, computed from the proper times and areal radii recorded by the Lagrangian tracers.}
\label{Fig:5}
\end{figure}

In Fig.~\ref{Fig:5} we show a comparison between our numerical results and the analytical diffusion approximation.  For these simulations, we choose an initial areal radius $R_0 = 10 M$, we set up the initial data as described in \cite{StaBBFS12}, and we approximate dust as a fluid with a Gamma-law equation of state (\ref{gamma_law}) with $\Gamma = 1.001$ and with $P = 10^{-6} \rho_0$ initially.  We also choose the initial radiation energy density to be $E(0) = E_0 = 10^{-5} \rho_0$, and impose local thermal equilibrium by setting $B = E / (4 \pi)$ as in the analytical solution of \cite{Sha89}, and set the initial flux $F^a$ to zero.  With these choices the pressure is radiation dominated and has little influence on the dynamics ($P / \rho_0 \ll M/r$).  We adopted $N_r = 2048$ radial equidistant grid points extending to an exterior outer boundary at an isotropic radius of $r_{\rm max} = 24 M$, and evolved with moving-puncture coordinates, i.e.~1+log slicing for the lapse \cite{BonMSS95}, and a Gamma-driver condition for the shift \cite{ThiBB11,ThiBHBR11}.  In the notation of Eq.~(39) in \cite{StaBBFS12} we chose the parameter $\mu_S$ appearing in the Gamma-driver condition according to $\mu_S = \alpha^2$.   

Since we restrict our analysis to the optically thick stellar interior, we impose radiation boundary conditions close to the stellar surface.  For strictly outgoing isotropic emission at the stellar surface the boundary condition is $F = 0.5 E$, where $F$ is the magnitude of the flux 
\begin{equation} \label{flux_magnitude}
F \equiv \left( F_a F^a \right)^{1/2} = \left( \Fs_i \Fs^i - \Fs^2 \right)^{1/2}.
\end{equation}
Since $E$ quickly plummets at the surface once the evolution is underway, we follow \cite{Sha89} and adopt the ``zero-temperature approximation" for $E$ at the surface, i.e.~$E=0$. The flux is much smaller than $E$ everywhere in the interior and we find that its behavior is insensitive to its precise boundary value we choose near and at the surface provided it is kept small. In keeping with our zero-temperature approximation for $E$ we therefore set $\Fs^i = 0$ near the surface.  We caution that our closure relation (\ref{closure}) does not provide a realistic prescription in the optically thin regions very close to the surface, but in the limit of arbitrarily large values of the opacity this region is infinitesimally thin geometrically.  We have chosen the absorption opacity $\kappaa$ so that the optical depth of the center is $\tau^{\rm abs} = \kappaa \rho_0 R = 25$, as well as $\kappas = 0$.  Note that the optical depth increases as $R^{-2}$ as the collapse proceeds.

We follow 25 Lagrangian fluid tracers, and record the fluid and radiation variables observed by these fluid particles together with their proper times $\tau$ and areal radii $R$.  At selected instants of coordinate time $t$ we then plot these variables, and compare with the analytical solutions computed from $\tau$ and $R$.  Both the analytical and numerical solutions remain valid even after the entire star is inside a black hole.

We find very good agreement of the our numerical results with the analytical expression (\ref{rho_0_OS}) for the rest-mass density $\rho_0$ (top panel in Fig.~\ref{Fig:5}), and -- consistent with the findings of \cite{Sha96} -- quite good agreement with the diffusion result for the radiation energy density $E$ (middle panel in Fig.~\ref{Fig:5}).  Likewise, the magnitude of the flux $F$ (lower panel), which, unlike the components of $F^a$, is a scalar and can be compared directly, is in reasonably good agreement after the initial transient.  Similar to the findings of \cite{Sha96}, the values on the surface, which determine the emergent flux, are not all that different from the diffusion values, but during the initial transient the behavior is quite different in the stellar interior.  The numerical data for the flux drop to very small values probably dominated by numerical truncation error even close to the surface, while the analytical solution follows an approximately exponential decay toward greater optical depths.  In order to better understand these differences, and to illuminate some of the features of the numerical solution, we analyze the behavior of solutions to the dynamical radiation equations (\ref{tau_dot}) and (\ref{S_dot}) at early times.

\subsubsection{Initial transient -- analytical treatment}
\label{sec:numerics_OS_early_analytical}

We will assume again that neither the spacetime nor the dust evolution are affected by the radiation field, so that both are still given by the expressions of Section \ref{sec:numerics_OS_OS}.  Adopting the same Gaussian normal coordinates as used there we can identify the lapse $\alpha = 1$, the shift vector $\beta^i = 0$, and the spatial metric
\begin{equation}
\gamma_{ij} = a^2(\tau) 
\left(
\begin{array}{ccc}
1 & 0 & 0 \\
0 & \sin^2 \chi & 0 \\
0 & 0  & \sin^2 \chi \, \sin^2 \theta 
\end{array}
\right)
\end{equation}
from the line element (\ref{OS_g}).  In these coordinates, slices of constant coordinate time $t$ coincide with slices of constant proper timer $\tau$, and the 
mean curvature on these slices is given by
\begin{equation} \label{K}
K \equiv \gamma^{ij} K_{ij} = - \frac{3}{a} \, \frac{da}{d\tau}.
\end{equation}
We also note that, in these comoving coordinates, both normal observers and dust particles follow geodesics; therefore the normal vector $n^a$ on slices of constant $\tau$ must be aligned with the dust's four-velocity $u^a$, $n^a = u^a$.   From (\ref{v_def}) we see that the dust's spatial velocity $v^a$ must therefore vanish in these coordinates, $v^a = 0$, and that $W = - n_a u^a = 1$.  

In order to evaluate the dynamical equations (\ref{tau_dot}) and (\ref{S_dot}) for heated Oppenheimer-Snyder collapse we first choose 
\begin{equation} \label{reference_OS}
\hat \gamma_{ij} = 
\left(
\begin{array}{ccc}
1 & 0 & 0 \\
0 & \sin^2 \chi & 0 \\
0 & 0  & \sin^2 \chi \, \sin^2 \theta 
\end{array}
\right)
\end{equation}
as our reference metric, so that
\begin{equation}
\sqrt{ \gamma / \hat \gamma } = a^3.
\end{equation}
We note that this ratio depends on $\tau$ only; in particular it remains finite and non-zero at the center of the coordinate system, highlighting another advantage of the reference-metric formalism.  Without using this formalism we would have encountered the term $\gamma^{1/2} = a^3 \sin^2 \chi \, \sin \theta$ instead, which vanishes at the origin, and displays a significantly more complicated dependence on the coordinates.  Similarly, in spherical polar coordinate systems, $\gamma^{1/2}$ itself typically scales with $r^2 \sin \theta$ close to the origin.  In the reference-metric formalism, this term can be canceled out by choosing the reference metric $\hat \gamma_{ij}$ to be the flat metric in spherical polar coordinates -- thereby avoiding the numerical issues associated with a vanishing determinant.  This is essentially what we did in (\ref{reference_OS}). 

We also note that, from (\ref{u_dot_F_2}) with $v^a = 0$, we must have
\begin{equation}
\Fs = 0.
\end{equation}
We then have
\begin{equation}
\bar \tau = a^3 \bar \rho = a^3 E
\end{equation}
from (\ref{tau}) and (\ref{rhorad}), 
\begin{equation}
f_{\bar \tau}^i = a^3 \Fs^i
\end{equation}
from (\ref{f_tau}), and
\begin{equation}
s_{\bar \tau} = \frac{a^3}{3} E \gamma^{ij} K_{ij} = - a^2 E \frac{da}{d\tau}
\end{equation}
from (\ref{s_tau}), where we have used (\ref{K}) in the last expression.  Inserting the last three equations into the energy equation (\ref{tau_dot}) we obtain
\begin{equation} \label{tau_dot_OS}
\partial_\tau (a^4 E) + a \Dref_i (a^3 \Fs^i ) = 0,
\end{equation}
where we have used $d\tau = dt$ in Gaussian normal coordinates.  

We similarly evaluate (\ref{S}), (\ref{S_flux}) and (\ref{s_S}) to find
\begin{align}
\bar S_i & = a^3 \Fs_i, \label{S_Reg_I} \\
(f_{\bar S})_i^{~j} & = \frac{a^3}{3} E \, \delta_i^{~j}, \\
(s_{\bar S})_i & = \frac{a^3}{6} E \, \gamma^{jk} \Dref_i \gamma_{jk} = 0 \label{s_S_i}
\end{align}
(where we have used $\Dref_i \gamma_{jk} = \Dref_i (a^2 \hat \gamma_{jk}) = 0$ in the last expression), and insert these into the momentum equation (\ref{S_dot}) to find 
\begin{equation} \label{S_dot_OS}
\partial_\tau (a^3 \Fs_i) + \frac{1}{3} \Dref_j (a^3 E \delta_i^{~j} ) = - a^3 \rho_0 (\kappaa + \kappas) \Fs_i.
\end{equation}
We briefly note that the term $(s_{\rm \bar S})_i$ in (\ref{s_S_i}) vanishes by virtue of the reference-metric formalism; without this formalism, the covariant derivative $\Dref_i$ would appear as a partial derivative $\partial_i$ instead, and one would rely on the resulting non-zero terms to be canceled by new terms originating from the appearance of $\gamma^{1/2}$ rather than $(\gamma / \hat \gamma)^{1/2}$ in the divergence term on the left-hand side of (\ref{S_dot_OS}) (see also the discussion in Section III.E in MBM).

Equations (\ref{tau_dot_OS}) and (\ref{S_dot_OS}) now form a pair of equations for the two primitive variables $E$ and $\Fs^i$.  Combining the two equations, one can show that, as a consequence of our adopted closure relation (\ref{closure}), the characteristic speeds of the radiation field take the expected values $c_{\rm rad} = \pm \sqrt{1/3}$ as measured by a normal observer comoving with the matter.   In the schematic spacetime diagram of Fig.~\ref{Fig:OS} we include one such characteristic, originating at the stellar surface at the initial time and traveling towards the stellar center, as the (green) line labeled ``radiation char.".  

Now consider as initial data a homogeneous radiation energy density $E(0) = E_0$ throughout the star, and zero flux $\Fs^i(0) = 0$ everywhere.  By contrast, $E(0)$ is set equal to zero outside the star, which distinguishes the stellar surface.  For these data, the spatial derivatives in equations (\ref{tau_dot_OS}) and (\ref{S_dot_OS}) vanish identically in the interior.  In the domain of dependence of the interior initial data, the flux will then remain zero, $\Fs^i(\tau) = 0$, while the energy equation (\ref{tau_dot_OS}) is solved by adiabatic heating
\begin{equation} \label{E_OS_early}
\frac{E(\tau)}{E_0} = \left( \frac{a_m}{a(\tau)} \right)^4,
\end{equation}
as one might have expected.  In the spacetime diagram of Fig.~\ref{Fig:OS}, the domain of dependence of the interior initial data is given by the area marked as Regions I and II, below the radiation characteristic originating from the surface.  In these two regions, the analytical solution to the radiation equations is given by (\ref{E_OS_early}) together with $\Fs^i = 0$ {\em in any coordinate system}.  Only in Region III can the radiation field approach the diffusive analytical solution of \cite{Sha89,Sha96}.  This reflects the difference between the full transport equations, which are hyperbolic, and the diffusion approximation, which is parabolic.

\subsubsection{Initial transient -- numerical results}
\label{sec:numerics_OS_early_numerical}

\begin{figure}[t]
\includegraphics[width = 0.45 \textwidth]{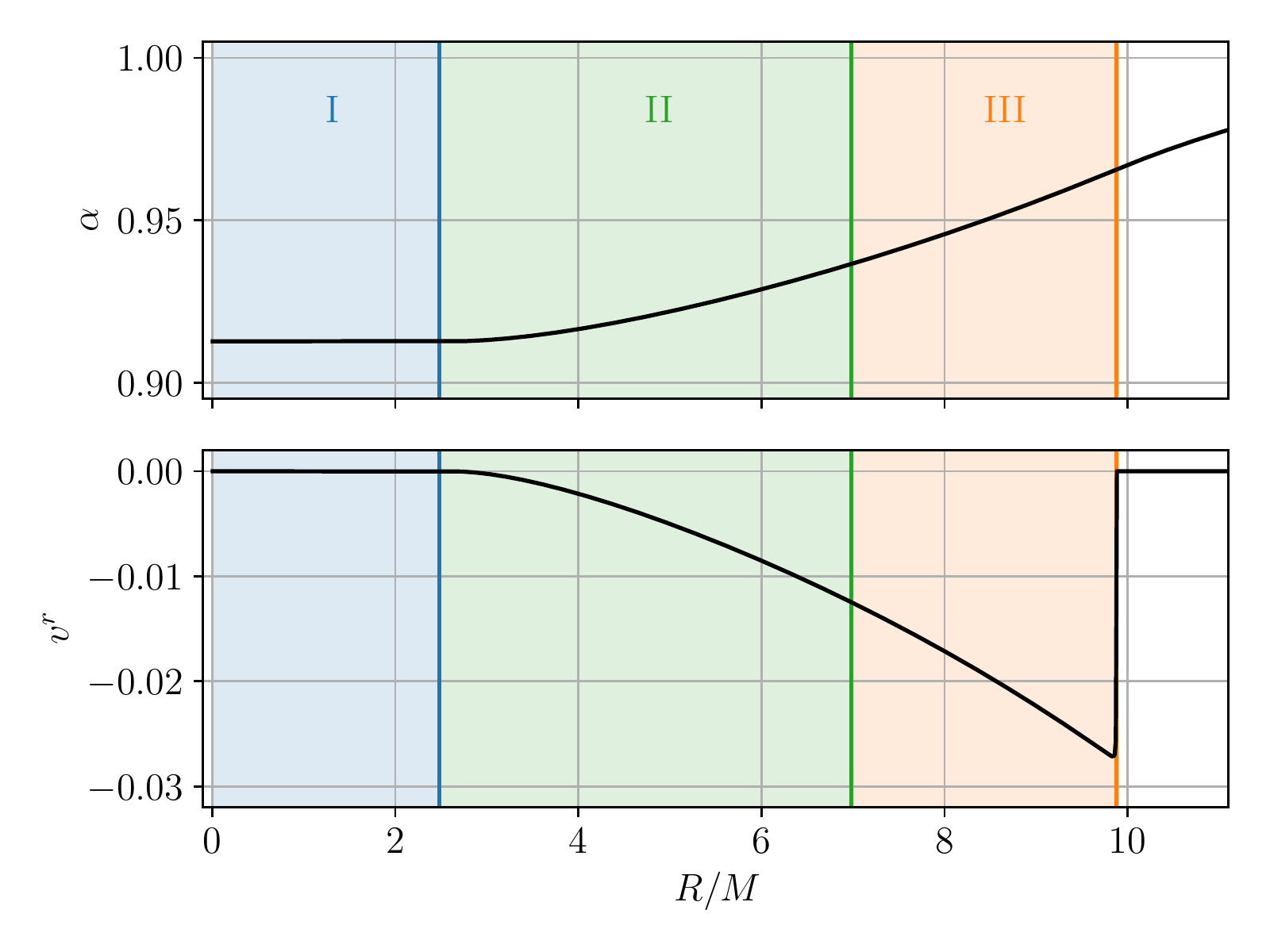}
\caption{The lapse $\alpha$ and the dust velocity $v^r$ at coordinate time $t = 5.52 M$ (compare the (green) circled data in Fig.~\ref{Fig:5}).  Regions I, II, and III are labeled as in the schematic spacetime diagram of Fig.~\ref{Fig:OS}.  Note that $\alpha$ does not depend on $R$, and $v^r = 0$, in Region I.}
\label{Fig:6}
\end{figure}

Numerical codes usually do not adopt Gaussian coordinates, however; instead, a common choice are moving-puncture coordinates with 1+log slicing \cite{BonMSS95},
\begin{equation} \label{1+log}
(\partial_t - \beta^i \partial_i) \alpha = - 2 \alpha K.
\end{equation}
The properties of Oppenheimer-Snyder collapse as rendered in 1+log slicing with initial condition $\alpha(0) = 1$ were analyzed by \cite{StaBBFS12}; in particular, the authors pointed out the existence of a gauge characteristic with characteristic speed $c_{\rm gauge} = \pm \sqrt{2 / \alpha}$ as measured by a normal observer.  In Fig.~\ref{Fig:OS}, the (blue) gauge characteristic labeled ``gauge char." originating from the surface at the initial time and propagating toward the center separates Region I from Region II.  As pointed by \cite{StaBBFS12}, the lapse remains spatially constant in Region I, and takes the value 
\begin{equation} \label{lapse}
\alpha = 1 + 6 \ln\left( a(\tau) / a_m \right)
\end{equation}
there, while outside of Region I the lapse will depend on space also.  In the top panel of Fig.~\ref{Fig:6} we show a snapshot at $t = 5.52 M$ (corresponding to the data shown as the (green) circles in Fig.~\ref{Fig:5}).  In Region I, where the lapse remains spatially constant, slices of constant coordinate time $t$ will coincide with slices of constant proper time $\tau$, as sketched in the schematic spacetime diagram of Fig.~\ref{Fig:OS}.  Outside of Region I, however, where the lapse is no longer spatially constant, slices of constant coordinate time depart from those of constant proper time (the dashed and dotted lines, respectively, in Fig.~\ref{Fig:OS}).  Furthermore, the normal vector $n^a$ is still be aligned with the dust's four-velocity $u^a$ in Region I, so that we still have $v^a = 0$ in this region, as shown in the bottom panel of Fig.~\ref{Fig:6}.  

\begin{figure}[t]
\includegraphics[width = 0.45 \textwidth]{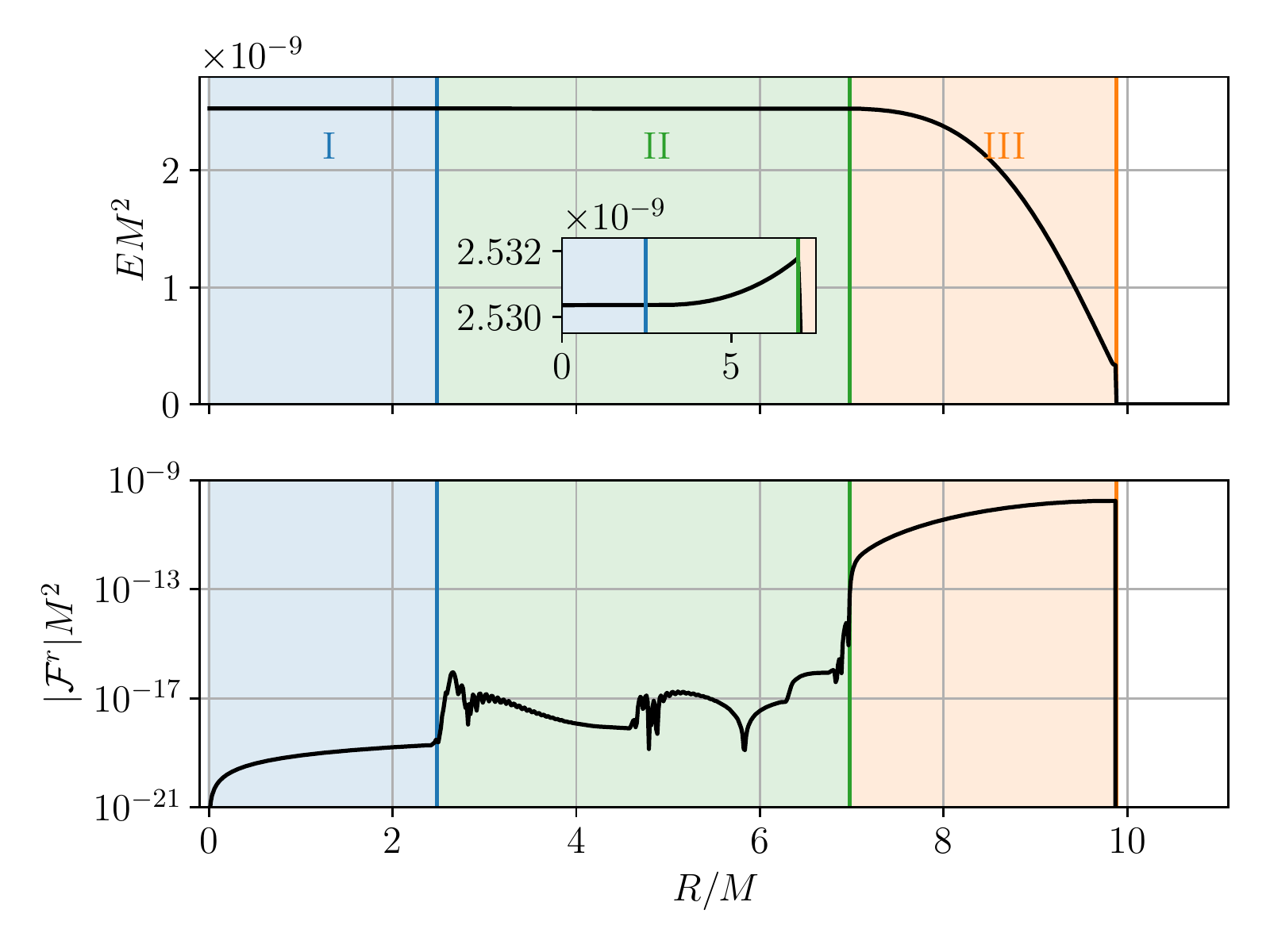}
\caption{Same as Fig.~\ref{Fig:6} but for the primitive radiation variables $E$ (top panel) and $\Fs^r$ (bottom panel).  The inset shows the small increase in $E$ towards larger radii in Region II.}
\label{Fig:7}
\end{figure}

\begin{figure}[t]
\includegraphics[width = 0.45 \textwidth]{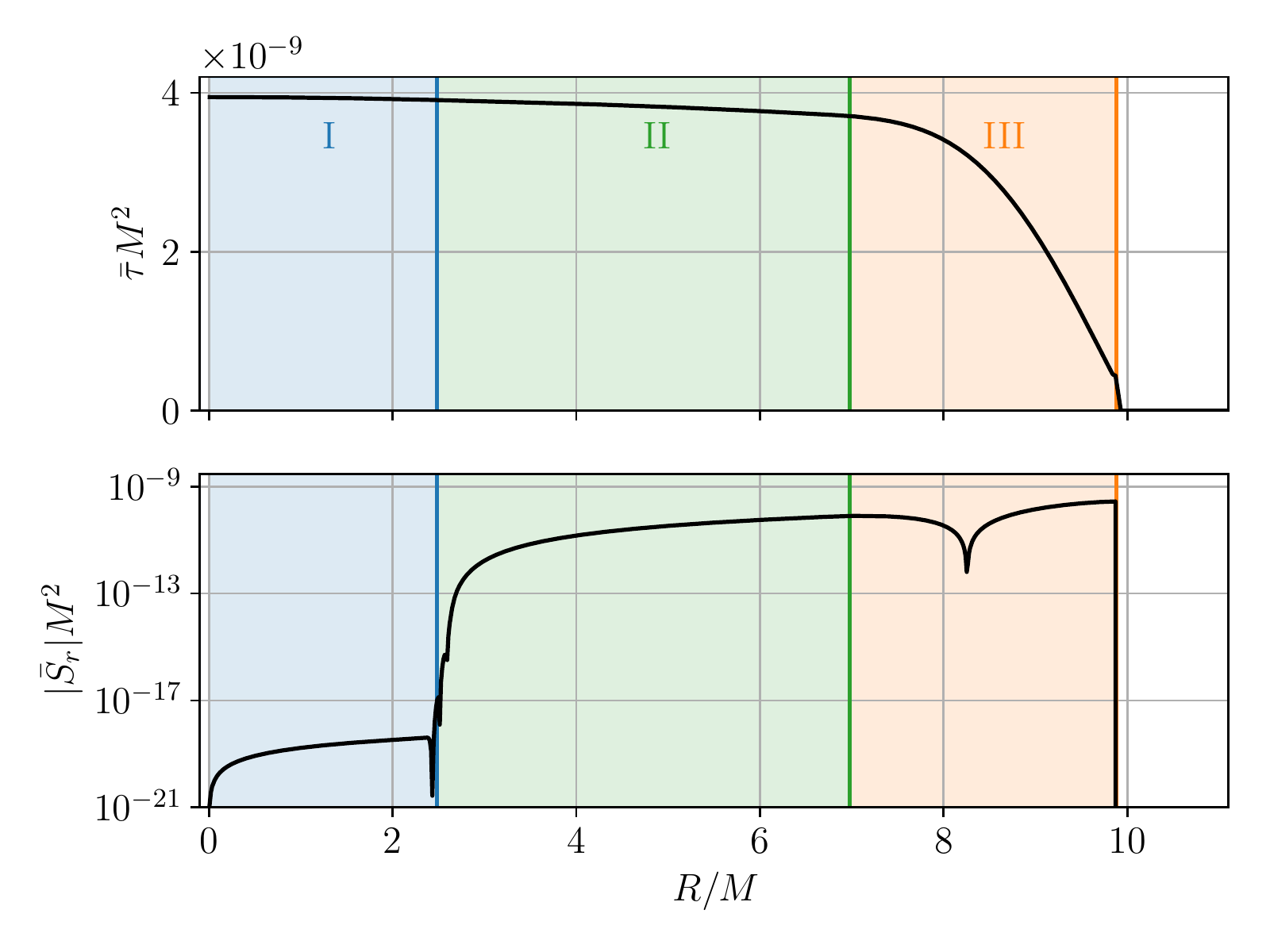}
\caption{Same as Fig.~\ref{Fig:6} but for the conserved radiation variables $\bar \tau$ (top panel) and $\bar S_r$ (bottom panel) in the stellar interior. }
\label{Fig:8}
\end{figure}

We can now discuss the consequences of these coordinate properties on the radiation quantities.  In Fig.~\ref{Fig:7} we show the primitive radiation energy density $E$ and flux $\Fs^r$.  As expected, $E$ is constant in Region I, according to (\ref{E_OS_early}) together with the observation that slices of constant $t$ and $\tau$ coincide in this region.  The latter is not the case in Region II; since a constant coordinate time $t$ corresponds to a later proper time $\tau$ at larger radius, the energy density $E$ slightly increases outwards in Region II (shown in the inset), before dropping significantly more rapidly in Region III, which has come into causal contact with the stellar surface.   We also see that the flux $\Fs^r$ is non-zero in Region III, but very close to zero in Regions I and II, as we would expect.  

Note, however, that $\Fs^r$ appears to be affected by significantly more numerical error in Region II than in Region I.  This behavior can be understood in terms of the conserved radiation quantities, shown in Fig.~\ref{Fig:8}.  As we discussed above, we have (up to numerical error from the hydrodynamical evolution and recovery) $v^r = 0$ in Region I (see bottom panel in Fig.~\ref{Fig:6}); we also expect $\Fs^r = 0$ and $\Fs = 0$ in this region.  According to equation (\ref{S_Reg_I}) this implies $\bar S_i = 0$ in Region I, consistent with our numerical results shown in the bottom panel of Fig.~\ref{Fig:8}.  We therefore expect that all terms on the right-hand side of equation (\ref{f_recovery}) will be small, and that we will hence be able to obtain the analytical solution $\Fs^i = 0$ to high accuracy.

Outside of Region I, however, the lapse is no longer spatially constant, which, as we discussed above, results in non-zero velocities $v^r$ (see the bottom panel of Fig.~\ref{Fig:6}).  By the same token, this results in non-zero values for $\bar S_i$ (see the bottom panel of Fig.~\ref{Fig:8}).   Solving the recovery equation (\ref{f_recovery}) in Region II, we see that we now compute a (vanishingly) small quantity $\Fs^i$ from differences between non-zero quantities; evidently, this will lead to larger numerical error than in Region I, as we observed in the bottom panel of Fig.~\ref{Fig:7}.  

Also note that $\bar S_r$ changes sign at around $R = 8.25 M$ in Region III; in the outer part we have $\bar S_r > 0$, reflecting an outward flux close to the surface, while in the inner parts, at larger optical depths, our normal observers see the radiation being dragged inward by the collapsing matter, so that $\bar S_r < 0$.

The behavior shown for $\Fs^r$ in Fig.~\ref{Fig:7} can also be seen for $F = F_a F^a$ at early times in Fig.~\ref{Fig:5}.   For $t = 0.92 M$ and $t = 5.52 M$, we can clearly distinguish Regions I, II and III.  At later times, following the initial transient, both the gauge and radiation characteristics have reached the center, the entire star is now in Region III, and the radiation solution starts to approximate closely that described by the solution to the diffusion equation.  The diffusion approximation, in turn, agrees quite well with the exact numerical solution of the Boltzmann equation \cite{Sha96} after the initial transition, even when exact boundary conditions are incorporated at the stellar surface.  

We point out that the heated Oppenheimer-Snyder collapse problem we probed here is specifically designed to highlight the difference between an exact hyperbolic and an approximate radiation diffusion (parabolic) treatment.  In particular, by adopting a very compact initial configuration ($R_0 / M = 10$) and matter that undergoes free-fall collapse at nearly the speed of light, the transient phase, during which the two approaches differ, takes up a non-negligible fraction of the total collapse time.  For more realistic scenarios the transient phase, which only lasts a few light travel times across the initial star, represents an insignificant fraction of the total evolution and radiative transport time.

\section{Summary}
\label{sec:summary}

We adopt a two-moment approximation together with a reference-metric formalism to bring the moment equations of relativistic radiation transfer into a form that is well suited for numerical implementations in curvilinear coordinates.  While curvilinear coordinates can be very effective in taking advantage of either exact or approximate symmetries, they also introduce coordinate singularities that can be problematic in numerical implementations.  One approach is to treat all singular terms analytically, and the reference-metric formalism provides a framework that allows such a treatment.  In this paper we derive the equations governing the radiation fields within this formalism, resulting in \eq~(\ref{tau_dot}) for the radiation energy density and \eq~(\ref{S_dot}) for the radiation momentum density, or flux.  In contrast to many previous treatments we also employ a systematic 3+1 decomposition of the radiation fields.  We focus here on the optically-thick regime and adopt an Eddington factor of 1/3 (see \eq~\ref{closure}), together with a gray (frequency-independent) opacity, but both restrictions can be relaxed.

The equations for the radiation fields take a form that is very similar to the corresponding equations of hydrodynamics; an existing relativistic hydrodynamics code can therefore be augmented to treat radiation as well by incorporating the radiation equations into the hydrodynamics algorithm.  We implement these equations in a code that adopts spherical polar coordinates, and, as numerical demonstrations, present results for two test problems.  Specifically, we consider stationary planar shock solutions in flat spacetimes, for which a semi-analytical solution can be constructed by solving ordinary differential equations, as well as heated Oppenheimer-Snyder collapse, for which we carefully analyze the transition from an early transient to a post-transient phase that is well approximated by an analytical-known relativistic diffusion solution.

Many astrophysical objects and processes -- including single stars, remnants of neutron star merger or supernova collapse, and accretion onto black holes -- display at least an approximate symmetry.  Taking advantage of these symmetries as effectively as possible usually entails adopting curvilinear coordinates, for example spherical polar or cylindrical coordinates.   Even when the matter fields lack symmetry, the radiation propagates radially at large distances, where it is measured.  The formalism presented in this paper provides one approach for such simulations, and we hope that it will prove useful for the modeling of radiation transport (EM and/or neutrinos) in a number of interesting and important astrophysics scenarios, including the above.

\acknowledgments

It is a pleasure to thank Yuk Tung Liu for helpful conversations.  This work was supported by National Science Foundation (NSF) grants PHY-1707526 and PHY-2010394 to Bowdoin College, NSF grants PHY-166221 and PHY-2006066 and National Aeronautics and Space Administration (NASA) grant 80NSSC17K0070 to the University of Illinois at Urbana-Champaign, and through sabbatical support from the Simons Foundation (Grant No.~561147 to TWB).


\end{document}